\documentclass{article}

\newif\ifdraft \drafttrue
 \draftfalse

\usepackage[toc,page,header]{appendix}
\usepackage{minitoc}

\PassOptionsToPackage{authoryear}{natbib}

\usepackage[final]{neurips_2021}

\usepackage[utf8]{inputenc} 
\usepackage[T1]{fontenc}    
\usepackage[colorlinks,citecolor=blue,urlcolor=blue,linkcolor=blue,linktocpage=true, bookmarks=false]{hyperref}       
\usepackage{url}            
\usepackage{booktabs}       
\usepackage{amsfonts}       
\usepackage{nicefrac}       
\usepackage{microtype}      
\usepackage{xcolor}         

\usepackage{subfiles}
\usepackage{amsthm}
\usepackage{amsmath}
\usepackage{bm}
\usepackage{csquotes}
\usepackage{wrapfig}
\MakeOuterQuote{"}

\usepackage{comment}

\usepackage{epsfig}
\usepackage{graphicx}
\usepackage{wrapfig}
\usepackage{subcaption}
\usepackage{multi-row}
\usepackage{graphicx}
\usepackage{caption}
\usepackage{array}
\usepackage{bbm}
\usepackage{color}
\usepackage{enumerate}
\usepackage{enumitem}
\setlist{leftmargin=10mm}
\usepackage{mathtools}
\usepackage{amsmath,amssymb,amsthm,bm}
\usepackage{amsfonts,graphicx}
\usepackage{mathrsfs}
\usepackage{algorithmic,algorithm}

\usepackage[american]{babel}

\DeclareMathOperator*{\argmin}{arg\,min}
\newtheorem{theorem}{Theorem}
\newtheorem{lemma}[theorem]{Lemma}
\newtheorem{corollary}[theorem]{Corollary}
\newtheorem{example}[theorem]{Example}
\newtheorem{proposition}[theorem]{Proposition}
\theoremstyle{definition}
\newtheorem{defn}[theorem]{Definition}

\newcommand\norm[1]{\left\lVert#1\right\rVert}
\renewcommand{\argmin}{\mathop{\mathrm{argmin}}}

\newcommand{\red}[1]{\textcolor{red}{#1}}
\newcommand{\blue}[1]{\textcolor{blue}{#1}}

\def\E{\mathbb{E}}
\def\P{\mathbb{P}}

\def \E{\mathbb{E}}
\def\R{\mathbb{R}}
\def\cA{\mathcal{A}}

\def\cN{\mathcal{N}}

\def\cS{\mathcal{S}}

\def\cX{\mathcal{X}}
\def\cY{\mathcal{Y}}
\def\cZ{\mathcal{Z}}

\newcommand{\yw}[1]{\ifdraft\textit{\textcolor{red}{[yuxiang]: #1}}\fi} 

\author{%
  Rachel Redberg \\
  Department of Computer Science\\
  UC Santa Barbara\\
  Santa Barbara, CA 93106 \\
  \texttt{rredberg@ucsb.edu} \\
   \And
   Yu-Xiang Wang \\
   Department of Computer Science \\
   UC Santa Barbara\\
   Santa Barbara, CA 93106 \\
   \texttt{yuxiangw@cs.ucsb.edu} \\
}

\title{Privately Publishable Per-instance Privacy}

\begin{document}

\doparttoc 
\faketableofcontents 
\maketitle


    \begin{abstract}
We consider how to privately share the personalized privacy losses incurred by objective perturbation, using per-instance differential privacy (pDP). Standard differential privacy (DP) gives us a worst-case bound that might be orders of magnitude larger than the privacy loss to a particular individual relative to a fixed dataset. The pDP framework provides a more fine-grained analysis of the privacy guarantee to a target individual, but the per-instance privacy loss itself might be a function of sensitive data. In this paper, we analyze the per-instance privacy loss of releasing a private empirical risk minimizer learned via objective perturbation, and propose a group of methods to privately and accurately publish the pDP losses at little to no additional privacy cost.    \end{abstract}

\section{Introduction}
\label{introduction}

An explosion of data has fueled innovation in machine learning applications and demanded, in equal turn, privacy protection for the sensitive data with which machine learning practitioners train and evaluate models.

Differential privacy (DP) \citep{dwork2006calibrating, dwork2014algorithmic} has become a mainstay of privacy-preserving data analysis, replacing less robust privacy definitions such as \textit{k}-anonymity which fail to protect against sufficiently powerful de-anonymization attacks \citep{narayanan2008robust}. In contrast, DP offers provable privacy guarantees that are robust against an arbitrarily strong adversary.

The data curator could trivially protect against privacy loss by reporting a constant function, or by releasing only data-independent noise. The key challenge of DP is to release privatized output that retains utility to the data analyst.

A desired level of utility in a machine learning application might necessitate a high value of $\epsilon$, but the privacy guarantees degrade quickly past $\epsilon = 1$. \citep{triastcyn2020bayesian} construct an example whereby a differentially private algorithm with $\epsilon = 2$ allows an attacker to use a maximum-likelihood estimate to conclude with up to 88\% accuracy that an individual is in a dataset. For $\epsilon = 5$, the theoretical upper bound on the accuracy of an optimal attack is 99.3\%.

Moreover, practical applications of differential privacy commonly use large values of $\epsilon$. A study of Apple's deployment of differential privacy revealed that the overall daily privacy loss permitted by the system was as high as $\epsilon = 6$ for Mac OS 10.12.3 and $\epsilon = 14$ for iOS 10.1.1 \citep{tang2017privacy} -- offering only scant privacy protection!

Recent work \citep{yu2021large} has empirically justified large privacy parameters by conducting membership inference attacks to demonstrate that these seemingly tenuous privacy guarantees are actually much stronger in practice. These results are unsurprising from the perspective that DP gives a data-independent bound on the worst-case privacy loss which is likely to be a conservative estimate of the risk to a particular individual when a DP algorithm is applied to a particular input dataset.

\emph{Per-instance differential privacy} provides a theoretically sound alternative to the empirical approach for revealing the gap between the worst-case DP bound and the actual privacy loss in practice. The privacy loss to a particular individual relative to a fixed dataset might be orders of magnitude smaller than the worst-case bound guaranteed by standard DP. In this case, an algorithm meeting a desired level of utility but providing weak DP guarantees may, for the same level of utility, achieve drastically more favorable $\emph{per-instance}$ DP guarantees.

The remaining challenge is that the per-instance privacy loss is a function of the entire dataset; publishing it directly would negate the purpose of privately training a model in the first place! In this paper, we propose a methodology to privately release the per-instance privacy losses associated with private empirical risk minimization. Our contributions are as follows:
\begin{itemize}
    \item We introduce \emph{ex-post} per-instance differential privacy to provide a sharp characterization of the privacy loss to a particular individual that adapts to both the input dataset and the algorithm's output.
    \item We present a novel analysis of the \emph{ex-post} per-instance privacy losses incurred by the objective perturbation mechanism, demonstrating that these \emph{ex-post} pDP losses are orders of magnitude smaller than the worst-case guarantee of differential privacy.
    \item We propose a group of methods to privately and accurately release the \emph{ex-post} pDP losses. In the particular case of generalized linear models, we show that we can accurately publish the private \emph{ex-post} pDP losses using a dimension- and dataset-independent bound.
    \item One technical result of independent interest is a new DP mechanism that releases the Hessian matrix by adding a \emph{Gaussian Orthogonal Ensemble} matrix, which improves the classical ``AnalyzeGauss'' \citep{dwork2014analyze} by roughly a constant factor of $2$.
\end{itemize}


\subsection{Related Work}
\label{related_work}

This paper builds upon \citep{wang2019per}, which proposed the per-instance DP framework and left as an open question the matter of publishing the pDP losses. We extend the pDP framework to an \emph{ex-post} setting to provide privacy guarantees that adapt even more fluidly to data-dependent properties of our algorithms. Another fundamental ingredient in our privacy analysis is the objective perturbation algorithm (\texttt{Obj-Pert}) of \citep{chaudhuri2011differentially}, further analyzed by \citep{kifer2012private}, which privately releases the minimizer of an empirical risk by adding a linear perturbation to the objective function before optimizing. 

Per-instance DP and \emph{ex-post} per-instance DP belong to a growing family of DP definitions that provide a more fine-grained characterization of the privacy loss. Among these are data-dependent DP \citep{papernot2018}, which conditions on a fixed dataset; personalized DP \citep{ghosh2011selling, ebadi2015differential, liu2015fast}, which conditions on a fixed individual’s datapoint; and \emph{ex-post} DP \citep{ligett2017accuracy}, which conditions on the realized output of the algorithm. Per-instance DP conditions on both a fixed dataset and a fixed individual’s datapoint, and \emph{ex-post} per-instance DP adapts even further to the realized output of the algorithm. A more detailed comparison of these DP variants is included in the supplementary materials.

Other data-adaptive methodologies include propose-test-release \citep{dwork2009differential} and local sensitivity \citep{nissim2007smooth}. In addition, Bayesian differential privacy \citep{triastcyn2020bayesian} provides data-dependent privacy guarantees that afford strong protection to "typical" data by making distributional assumptions about the sensitive data. The R\'enyi-DP-based privacy filters of \citep{feldman2020individual} are also closely related to our work; the authors study composition of personalized (but not per-instance) privacy losses using adaptively-chosen privacy parameters.

\section{Preliminaries}

\subsection{Symbols and Notation}
\label{notation}

We write the output of a randomized algorithm $\mathcal{A}$ as $\mathcal{A}(\cdot)$, and for continuous distributions we take $\text{Pr}[\mathcal{A}(D) = o]$ to be the value of the probability density function at output $o$.

We will let $z\in \cZ$ refer to both an individual and their data; for example, individual $z$ holds data $z=(x, y)$ in a supervised learning problem. We take $\cZ^*=\cup_{n=0}^\infty \cZ^n$ to be the space of datasets with an unspecified number of data points. $D_{\pm z} \in \mathcal{Z}^*$ denotes the fixed dataset $D = \{z_1,\ldots,z_n\} \in \mathcal{Z}^*$ with the data point $z$ removed from $D$ if $z \in D$, 
or added to $D$ if $z \notin D$. 
In our mathematical expressions, we use "$\pm$" to mean "add if $z \notin D$, subtract otherwise". Similarly. "$\mp$" means "subtract if $z \notin D$, add otherwise".

We distinguish between $\epsilon$ as fixed input to a DP algorithm, and $\epsilon(\cdot)$ as a function parameterized according to a particular DP relaxation ---  e.g., $\epsilon(o, D, D_{\pm z})$ means the \textit{ex-post} per-instance privacy loss conditioned on output $o$, dataset $D$, and data point $z$.

\subsection{Differential Privacy}
\label{dp}

    Let $\mathcal{Z}$ denote the data domain, and $\mathcal{R}$ the set of all possible outcomes of algorithm $\mathcal{A}$. Fix $\epsilon, \delta \geq 0$.
    \begin{defn} (Differential privacy)  \label{def:dp}
        A randomized algorithm $\mathcal{A}: \mathcal{Z}^* \rightarrow \mathcal{R}$ satisfies $(\epsilon, \delta)$-DP if for all datasets $D \in \mathcal{Z}^{*}$ and data points $z \in \mathcal{Z}$, and for all measurable sets $S \subset \mathcal{R}$, 
        \begin{align*}
            &\text{Pr}\big[\mathcal{A}(D) \in S\big] \leq e^{\epsilon}\text{Pr}\big[\mathcal{A}(D_{\pm z}) \in S\big] + \delta.
        \end{align*}
    \end{defn}
    
    Differential privacy guarantees that the presence or absence of any particular data record has little impact on the output distribution of a randomized algorithm. In this paper we use the "add/remove" notion of DP, by which we construct neighboring dataset $D_{\pm z}$ by adding or removing an individual $z$ from dataset $D$. 
    
    DP is powerful and universal in that its guarantee applies to any $D, z$ and set of output events.  However, there are often situations where the privacy losses of $\mathcal{A}$ vary drastically depending on its input data, and the  privacy loss bound $\epsilon$ (protecting even the worst-case pair of neighboring datasets) may not be informative of the privacy loss incurred to individuals when the input to $\mathcal{A}$ is typical.  This motivated \citep{wang2019per} to consider a per-instance version of the DP definition. 
        \begin{defn} (Per-instance differential privacy)  \label{def:pdp}
    	A randomized algorithm $\mathcal{A}: \mathcal{Z}^* \rightarrow \mathcal{R}$ satisfies $\big(\epsilon(D, D_{\pm z}), \delta\big)$-pDP if for dataset $D$ and data point $z$, and for all measurable sets $S \subset \mathcal{R}$, 
    	\begin{align*}
    	&\text{Pr}\big[\mathcal{A}(D) \in S\big] \leq e^{\epsilon}\text{Pr}\big[\mathcal{A}(D_{\pm z}) \in S\big] + \delta,\\
    	&\text{Pr}\big[\mathcal{A}(D_{\pm z}) \in S\big] \leq e^{\epsilon}\text{Pr}\big[\mathcal{A}(D) \in S\big] + \delta.
    	\end{align*}
    \end{defn}
The pDP definition can be viewed as using a function $\epsilon(D,D_{\pm z})$ that more precisely describes the privacy guarantee in protecting a fixed data point $z$ when $\mathcal{A}$ is applied to dataset $D$. 

    
%
%


As it turns out, it is most convenient for us to work with an even more  \emph{instance-specific} description of the privacy loss that is further parameterized by the realized output of $\mathcal{A}$ \emph{ex-post} --- after the random coins of $\mathcal{A}$ are flipped and the outcome released.
\begin{defn} (\textit{Ex-post} per-instance differential privacy)\label{def:expost_pdp}
	A randomized algorithm $\mathcal{A}$ satisfies $\epsilon(\cdot)$-\emph{ex-post} per-instance differential privacy for an individual $z$ and a fixed dataset $D$ at an outcome $o\in \mathrm{Range}(\mathcal{A})$ if 
	$$
	\left|\log\left(\frac{\text{Pr}\big[\mathcal{A}(D) = o \big]}{\text{Pr}\big[\mathcal{A}(D_{\pm z})= o\big]}\right)\right| \leq \epsilon(o,D,D_{\pm z}). 
	$$ 
\end{defn}
This definition generalizes the \emph{ex-post} DP definition \citep{ligett2017accuracy} (introduced for a different purpose) to a \emph{per-instance} version that depends on a given pair of neighboring datasets.
The above quantity is essentially the absolute value of the log-odds ratio, used extensively in hypothesis testing. Intuitively, the $\emph{ex-post}$ per-instance privacy loss $\epsilon(o, D, D_{\pm z})$ describes how confidently an attacker could infer, given the output of algorithm $\mathcal{A}$, whether or not individual $z$ is in dataset $D$.


Despite (or perhaps because of) its precise accounting for privacy, \emph{ex-post} pDP could reveal sensitive information about the dataset, as the following example explicitly illustrates.
\begin{example}[The privacy risk of exposing \emph{ex-post} pDP]\label{ex:privacy_risk}
Consider a standard Gaussian mechanism $\mathcal{A}$ that adds noise to a counting query $Q$ applied to dataset $D$, i.e. $\mathcal{A}(D) = Q(D) + \mathcal{N}(0, \sigma^2)$. $Q$ has global sensitivity $\Delta_Q = 1$.  We will show that an attacker, knowing only the output $o$ of algorithm $\mathcal{A}$, her \emph{ex-post} pDP loss and that her individual data is not contained in dataset $D$, can conclusively uncover the sensitive quantity $Q(D)$ protected by algorithm $\mathcal{A}$.

Following the proof of Theorem \ref{pdp_gaussian}, the \emph{ex-post} pDP can be directly calculated as 
\begin{align*}\epsilon(o, D,D_{\pm z}) = \frac{|Q(D)-Q(D_{\pm z})||2o - Q(D) -  Q(D_{\pm z})| }{2\sigma^2}.\end{align*}

Enter attacker $z$, who has auxiliary information: she knows that her own individual data is not contained in $D$. After algorithm $\mathcal{A}$ is applied to $D$, attacker $z$ receives output $o = 1$ and is informed of her \emph{ex-post} pDP $\epsilon(o, D,D_{+z})$. Since $Q(D_{+z}) = Q(D) + 1$ is known, attacker $z$ can solve for $Q(D)$ and obtain $Q(D) = o - 0.5 \pm \sigma^2 \epsilon(o, D,D_{+z})$. With probability $1$, only one of the two possibilities is an integer\footnote{Take $Q(D) = 0$ and $o = 0.1$ as an example, the two possibilities are $0$ and $-0.8$.}. Therefore, exposing \emph{ex-post} pDP in this case completely reveals $Q(D)$.
\end{example}

\noindent\textbf{Problem statement.} The lesson of Example \ref{ex:privacy_risk} is that we cannot directly reveal the \emph{ex-post} pDP losses without potentially nullifying the algorithm's privacy benefits. How, then, can we privately and accurately publish the \emph{ex-post} pDP losses?

The goal of this paper is to develop an algorithm that publishes a \emph{function} $\tilde{\epsilon}: \mathcal{Z} \rightarrow \mathbb{R}$ whose output estimates the \emph{ex-post} pDP loss to an individual $z$ of releasing the output $\hat{\theta}^P$ from the objective perturbation mechanism. Any individual (not just those whose data is contained in the dataset) can plug her own data $z$ into this function in order to receive a high-probability bound on her \emph{ex-post} pDP loss which does not depend directly on any sensitive data except her own.

This requirement offers the same type of privacy protection as joint differential privacy \citep{kearns2014mechanism}, which relaxes the standard DP definition by allowing an algorithm's output to individual $z$ to be sensitive only in her own private data.  Our notion of privacy is slightly more general in that it holds for individuals both in and out of the dataset. The difference lies in how the algorithm's output space is defined; whereas a joint DP algorithm produces a fixed-length tuple partitioning the output to each individual in the dataset, our algorithm outputs a function whose domain includes any data point $z \in \mathcal{Z}$. As a result, our methods are robust against collusion by arbitrary coalitions of adversaries, allowing repeated queries by any group of individuals without invalidating the privacy guarantees promised by the pDP losses.


\subsection{Problem Setting}
\label{problem_setting}

We consider a general family of problems known as \emph{private empirical risk minimization} (ERM), which aim to approximate the solution to an ERM problem while preserving privacy. That is, we wish to privately solve optimization problems of the form
    \begin{align*}
        \hat{\theta} &= \argmin_{\theta \in \Theta}L(\theta ; D) + r(\theta),
    \end{align*}
    where $r(\theta)$ is a regularizer and $L(\theta ; D)= \sum_{i = 1}^n \ell(\theta; z_i)$ a loss function. Throughout, we assume that $\ell(\theta; z)$ and $r(\theta)$ are convex and  twice-differentiable with respect to $\theta$. Dataset $D$ is given by $D = \{ z_i\}_{i = 1}^n$, and $z_i = (x_i, y_i)$ for $x_i \in \mathcal{X} \subseteq \mathbb{R}^d$ and $y \in \mathcal{Y} \subseteq \mathbb{R}$, where $||x||_2 \leq 1$ and $|y| \leq 1$. We consider only unconstrained optimization over $\Theta = \mathbb{R}^d$.

    



\subsection{Objective Perturbation}
\label{sec:objective_perturbation}

    The objective perturbation algorithm solves
    \begin{align}
        \hat{\theta}^P &= \argmin\limits_{\theta \in \Theta} L(\theta ; D) + r(\theta) + \frac{\lambda}{2}||\theta||_2^2 + b^T\theta, \label{op}
    \end{align}
    where $b \sim \mathcal{N}(0, \sigma^2I_d)$ and parameters $\sigma, \lambda$ are chosen according to a desired $(\epsilon, \delta)$-DP guarantee.
    \begin{algorithm}[H]
        \caption{Release $\hat{\theta}^P$ via \texttt{Obj-Pert} \citep{kifer2012private}}
        \label{alg:Alg1}
        \begin{algorithmic}
            \STATE{{\bfseries Input:} Dataset $D$, noise parameter $\sigma$, regularization parameter $\lambda$, loss function $L(\theta; D) = \sum_i \ell(\theta; z_i)$, convex and twice-differentiable regularizer $r(\theta)$, convex set $\Theta$.}
            \STATE{{\bfseries Output:} $\hat{\theta}^P$, the minimizer of the perturbed objective.}
           \STATE{Draw noise vector $b \sim \mathcal{N}(0, \sigma^2I)$.}
           \STATE{Compute $\hat{\theta}^P$ according to (\ref{op}).}
        \end{algorithmic}
    \end{algorithm}
    
    \begin{theorem}[Privacy guarantees of Algorithm~\ref{alg:Alg1} \citep{kifer2012private}]
        \label{thm:alg1}
        Consider dataset $D = \{z_i\}_{i = 1}^n$; loss function $L(\theta; D) = \sum_i \ell(\theta; z_i)$; convex regularizer $r(\theta)$; and convex domain $\Theta$. Assume that $\nabla^2 \ell(\theta; z_i) \prec \beta I_d$ and $||\nabla \ell(\theta; z_i)||_2 \leq \xi$ for all $z_i \in \mathcal{X} \times \mathcal{Y}$ and for all $\theta \in \Theta$.
        For $\lambda \geq 2\beta/\epsilon_1$ and $\sigma = \xi^2(8 \log (2/\delta) + 4\epsilon_1)/\epsilon_1^2$, Algorithm~\ref{alg:Alg1} satisfies $(\epsilon_1, \delta)$-differential privacy.
    \end{theorem}
    
    The privacy guarantees stated in Theorem \ref{thm:alg1} apply even when $\theta$ is constrained to a closed convex set, but for ease of our per-instance privacy analysis we will require $\Theta = \mathbb{R}^d$ from this point on.

\section{Privately Publishable pDP}

\subsection{pDP Analysis of Objective Perturbation}
\label{pdp_analysis}

    Our goal in this section is to derive the personalized privacy losses (under Definition~\ref{def:expost_pdp}) associated with observing the output $\hat{\theta}^P$ of objective perturbation. This \emph{ex-post} perspective is highly adaptive and also convenient for our analysis of Algorithm~\ref{alg:Alg1}, whose privacy parameters are a function of the data. Since we are analyzing the per-instance privacy cost of \emph{releasing} $\hat{\theta}^P$, it makes perfect sense to condition the pDP loss on the privatized output of the computation.

Our first technical result is a precise calculation of the \emph{ex-post} pDP loss of objective perturbation.
    
    \begin{theorem}[\emph{ex-post} pDP loss of objective perturbation for a convex loss function]
\label{thm:pdp_alg1_general}

Let $J(\theta; D) = L(\theta ; D) + r(\theta) + \frac{\lambda}{2}||\theta||_2^2$ such that $L(\theta; D)  + r(\theta) = \sum_i \ell(\theta; z_i) + r(\theta)$ is a convex and twice-differentiable regularized loss function, and sample $b \sim \mathcal{N}(0, \sigma^2 I_d)$. Then for every privacy target $z = (x, y)$, releasing $\hat{\theta}^P = \argmin_{\theta \in \mathbb{R}^d} J(\theta; D) + b^T\theta$ satisfies $\epsilon_1(\hat{\theta}^P, D, D_{\pm z})$-\textit{ex-post} per-instance differential privacy with
\begin{align*}
    \epsilon_1(\hat{\theta}^P, D, D_{\pm z} ) =\left| -\log\prod\limits_{j = 1}^d \Big(1 \mp \mu_j \Big) + \frac{1}{2\sigma^2}||\nabla \ell(\hat{\theta}^P; z)||_2^2 \pm \frac{1}{\sigma^2} \nabla J(\hat{\theta}^P;D)^T\nabla \ell(\hat{\theta}^P; z)\right|,
\end{align*}
where $\mu_j = \lambda_j u_j^T\Big(\nabla b(\hat{\theta}^P; D) \mp \sum_{k = 1}^{j - 1} \lambda_k u_k u_k^T \Big)^{-1}u_j $ according to the eigendecomposition $\nabla^2 \ell(\theta; z) = \sum_{k = 1}^{d}\lambda_k u_ku_k^T$.
\end{theorem}
    
    \begin{proof}[Proof sketch]
    	Following the analysis of \citep{chaudhuri2011differentially}, we establish a bijection between the mechanism output $\hat{\theta}^P$ and the noise vector $b$, and use a change-of-variables defined by the Jacobian mapping between $\hat{\theta}^P$ and $b$ in order to rewrite the log-probability ratio in terms of the probability density function of $b$. First-order conditions then allow us to solve directly for the distribution of $b$. To calculate the first term of the above equation, we use the eigendecomposition of the Hessian $\nabla^2 \ell(\hat{\theta}^P; z)$ and recursively apply the matrix determinant lemma. The rest of the proof is straightforward algebra. The full proof is given in Appendix~\ref{app:proofs}.
    \end{proof}
    
    The above expression holds for any convex loss function, but is a bit unwieldy. The calculation becomes much simpler when we assume $\ell(\cdot)$ to be a generalized linear loss function, with inner-product form $\ell(\theta; z) = f(x^T \theta; y)$. For the sake of interpretability, we will defer further discussion of the \emph{ex-post} pDP loss of objective perturbation until after presenting the following corollary.
    
    \begin{corollary}[\emph{ex-post} pDP loss of objective perturbation for GLMs]
    \label{cor:pdp_alg1_linear}
            Let $J(\theta; D) = L(\theta ; D) + r(\theta) + \frac{\lambda}{2}||\theta||_2^2$ such that $L(\theta; D) = \sum_i \ell(\theta; z_i)$ is a linear loss function, and sample $b \sim \mathcal{N}(0, \sigma^2 I_d)$. Then for every privacy target $z = (x, y)$, releasing $\hat{\theta}^P = \argmin_{\theta \in \mathbb{R}^d} J(\theta; D) + b^T\theta$ satisfies $\epsilon_1(\hat{\theta}^P, D, D_{\pm z})$-\textit{ex-post} per-instance differential privacy with
        \begin{align*}
            \epsilon(\hat{\theta}^P, D, D_{\pm z}) &\leq
            \left|-\log \big(1 \pm f''(\cdot)\mu(x) \big) + \frac{1}{2\sigma^2}||\nabla \ell(\hat{\theta}^P; z)||_2^2  \pm  \frac{1}{\sigma^2} \nabla J(\hat{\theta}^P; D)^T\nabla \ell(\hat{\theta}^P; z) \right|,
        \end{align*}
        where $\mu(x) = x^T \big(\nabla^2 J(\hat{\theta}^P; D)\big)^{-1}x$, $\nabla \ell(\hat{\theta}^P; z) = f'(x^T\hat{\theta}^P; y)x$ and $f''(\cdot)$ is shorthand for $f''(\cdot) = f''(x^T \hat{\theta}^P; y)$. The notation $b(\hat{\theta}^P; D)$ means the realization of the noise vector $b$ for which the output of Algorithm \ref{alg:Alg1} will be $\hat{\theta}^P$ when the input dataset is $D$.
    \end{corollary}

 Note that the quantity $\mu(x)$ in the first term is the \emph{generalized leverage score} \citep{wei1998generalized}, quantifying the influence of a data point on the model fit. The second and third terms are a function of the gradient of the loss function and provide a complementary measure of how well the fitted model predicts individual $z$'s data.
\begin{figure}[H]
  \centering
    \includegraphics[width=\textwidth]{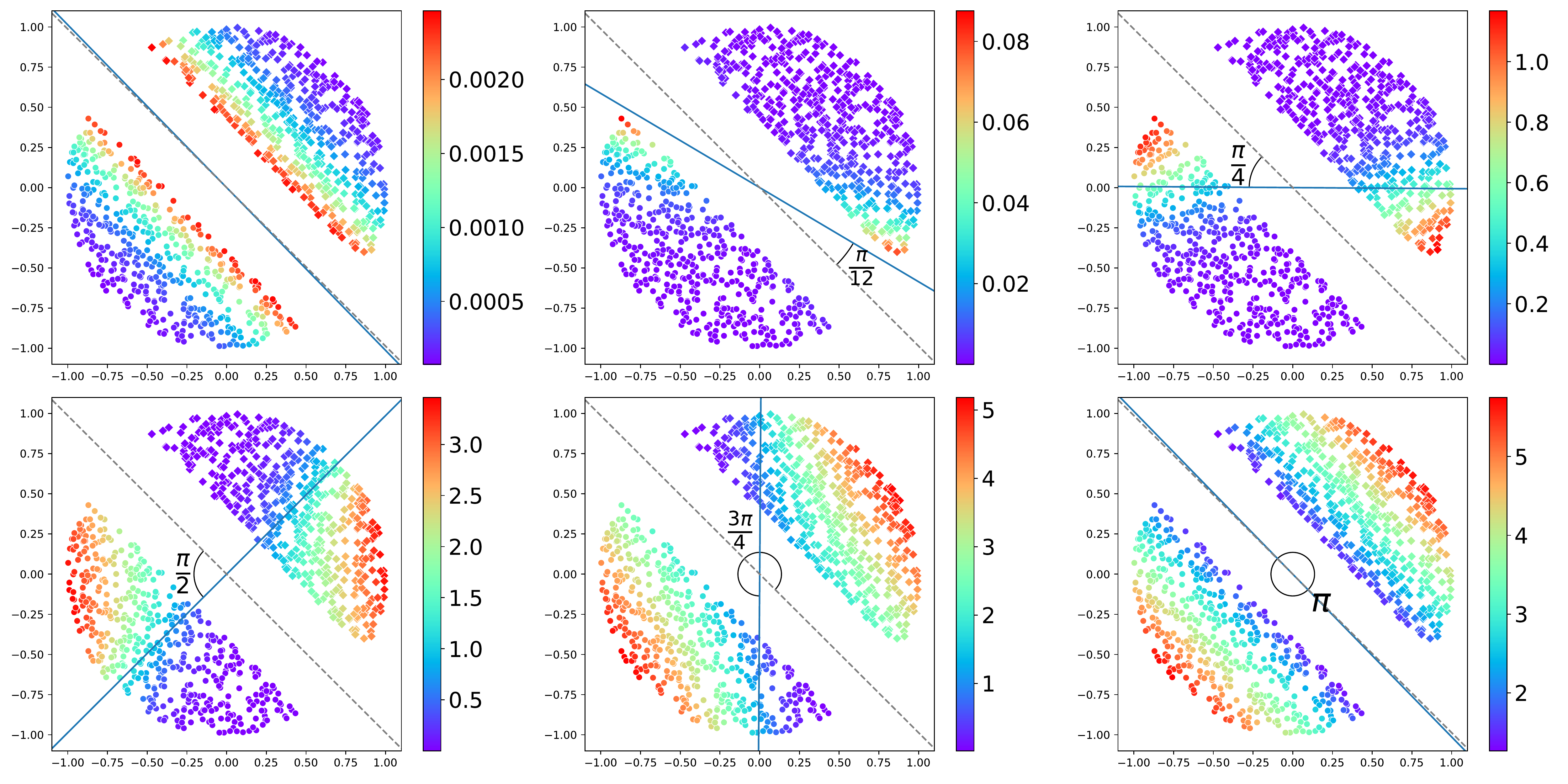}
  \caption{Visualization of \emph{ex-post} pDP losses for logistic regression ($n = 1000, d = 2$).}
  \label{fig:ex_post}
\end{figure}

Since the \emph{ex-post} pDP is a function of $\hat{\theta}^P$, we don't even need to run Algorithm ~\ref{alg:Alg1} to calculate \emph{ex-post} pDP losses -- we can plug in directly to Corollary ~\ref{cor:pdp_alg1_linear} in order to calculate the pDP distribution induced by any hypothetical $\hat{\theta}^P$. For Figure ~\ref{fig:ex_post}, we use a synthetic dataset $D$ sampled from the unit ball with two linearly separable classes separated by margin $m= 0.4$. Then we solve for $\hat{\theta} = \argmin J(\theta; D)$ with $\lambda = 1$ to minimize the logistic loss, and directly perturb the output by rotating it by angle $\omega \in [0, \frac{\pi}{12}, \frac{\pi}{4}, \frac{\pi}{2}, \frac{3\pi}{4}, \pi]$. We then denote $\hat{\theta}^P := \theta_{+ \omega}$ to mean $\theta$ rotated counter-clockwise by angle $\omega$. The color scale is a function of the \emph{ex-post} pDP loss of data point $z$. 

Figure ~\ref{fig:ex_post} illustrates how the mechanism output $\hat{\theta}^P$ affects the \emph{ex-post} pDP distribution of objective perturbation for our logistic regression problem. For $\omega \in [0, \frac{\pi}{12}]$, the data points closest to the decision boundary have the highest \emph{ex-post} pDP loss. These data points have a strong effect on the learned model and would therefore have high \emph{leverage scores}, making the first term dominate. As the perturbation (and model error) increases, the second and third terms dominate; the more badly a model predicts a data point, the less protection this data point has.

Hidden in this analysis are the $\delta$'s of Theorem ~\ref{thm:alg1}, which along with the choice of $\sigma$ and $\lambda$ could affect which of the three terms is dominant. Fortunately, the probability of outputting something like $\hat{\theta}^P = \theta_{+ \pi}$ is astronomically low for any reasonable privacy setting!

\subsection{Releasing the pDP losses}
\label{release_pdp}

    Next we consider: after having released $\hat{\theta}^P$ and calculated the per-instance privacy losses of doing so, how do we privately release these pDP losses? Our goal is to allow any individual $z\in \cZ$ (in the dataset or not) to know her privacy loss while preserving the privacy of others in the dataset.

   Observe that the expression from Theorem~\ref{thm:pdp_alg1_general} depends on the dataset $D$ only through two quantities: the leverage score $\mu(x) = x^T \big(\nabla^2 J(\hat{\theta}^P; D)\big)^{-1}x$ and the inner product $\nabla J(\hat{\theta}^P; D)^T \nabla \ell(\hat{\theta}^P; z)$. As a result, if we can find a data-independent bound for these two terms, or privately release them with only a small additional privacy cost, then we are done.
   
   \subsubsection{Data-independent bound of \emph{ex-post} pDP losses}
   \label{sub:data_indep_bounds}
   
   Below, we present a pair of lemmas which will allow us to find a high-probability, data-independent bound on the \emph{ex-post} pDP loss.
   
   \begin{theorem}
   \label{thm:data_indep_first}
   Suppose $\ell(\cdot)$ is a function with continuous second-order partial derivatives. Then
   \begin{align*}
       \left| -\log\prod\limits_{j = 1}^d \Big(1 \mp \mu_j \Big) \right| \leq -\sum_{j=1}^d\log(1 - \frac{\lambda_j}{\lambda}),
   \end{align*}
   where $\mu_j = \lambda_j u_j^T\Big(\nabla \mathbf{b}(\hat{\theta}^P; D) \mp \sum_{k = 1}^{j - 1} \lambda_k u_k u_k^T \Big)^{-1}u_j$ according to the eigendecomposition $\nabla^2 \ell(\hat{\theta}^P; z) = \sum_{k = 1}^{d}\lambda_k u_ku_k^T$. When specializing to linear loss functions such that $\ell(\theta; z) = f(x^T \theta; y)$, $\lambda_j=0$ for all $j>1$ and the above bound can be simplified to $-\log\left(1 - f''(x^T\hat{\theta}^P; y)||x||_2^2/\lambda\right)$.
   \end{theorem}
   
   \begin{theorem}\label{thm:data_indep_third}
   Let $\hat{\theta}^P$ be a random variable such that $\hat{\theta}^P = \argmin \left(J(\theta; D) + b^T \theta \right)$ as in (\ref{op}), where $b \sim \mathcal{N}(0, \sigma^2I_d)$ and $\ell(\theta; z)$ is a convex and twice-differentiable loss function. Then for $z \in \mathcal{Z}$, the following holds with probability $1 - \rho$:
   \begin{align*}
       \left|\nabla J(\hat{\theta}^P; D)^T \nabla \ell(\hat{\theta}^P; z)\right| \leq \sigma \sqrt{2 \log(2d/\rho)}\|\nabla \ell(\hat{\theta}^P; z)\|_1.
   \end{align*}
     For linear loss functions the bound can be substantially strengthened to
     \begin{align*}
         \left|\nabla J(\hat{\theta}^P; D)^T\nabla \ell(\hat{\theta}^P; z) \right| \leq f'(x^T\hat{\theta}^p;y)\sigma ||x||_2 \sqrt{2 \log (2/\rho)}.
     \end{align*}
   \end{theorem}
We make a few observations on the bounds. First, the general bound in Theorem~\ref{thm:data_indep_third} holds simultaneously for all $z$ and it depends only logarithmically in dimension when the features are \emph{sparse}. Second, the bound for a linear loss function is dimension-free and somewhat surprising because we are actually bounding an inner product of two \emph{dependent} random vectors (both depend on $\hat{\theta}^P$).
    
Finally, we remark that the bounds in this section are data-independent in that they do not depend on the rest of the dataset beyond already released information $\hat{\theta}^P$. It allows us to reveal a pDP bound of each individual when she plugs in her own data without costing any additional privacy budget! 

\subsection{The privacy report}
\label{privacy_report}

For certain regimes, we may wish to consider privatizing the data-dependent quantities of the \emph{ex-post} pDP losses, at an additional privacy cost, as an alternative to using data-independent bounds. Of course, it only makes sense to do so if we can show that (a) these data-dependent estimates are more accurate than the data-independent bounds; (b) the overhead of releasing additional quantities (the additional privacy cost in terms of both DP and pDP) is not too large; and (c) we can share the pDP losses of the private reporting algorithm using data-independent bounds (so we do not have to recursively publish such reports).

Full details are in the appendix. We show that by adding slightly more regularization than required by \texttt{Obj-Pert} (i.e., making $\lambda$ just a bit larger so that the minimum eigenvalue of the Hessian $H = \nabla^2 J$ is above a certain threshold), we can find a multiplicative bound that estimates $\mu(x) = x^T H^{-1} x$ uniformly for all $x$. We do so by adding noise to the Hessian using a natural variant of "Analyze Gauss" \citep{dwork2014analyze}, hence privately releasing $\overline{\mu^P}:\cX \rightarrow \R$. See Algorithm~\ref{alg:privacy_report} for details.  

For brevity, we use the short-hands $f'(\cdot):= f'(x^T \hat{\theta}^P; y)$ and $f''(\cdot) := f''(x^T\hat{\theta}^P; y)$, where $\ell(\theta; z) = f(x^T\theta; y)$ for GLMs. $F^{-1}_{\mathcal{N}(0, 1)}$ is the inverse CDF of the standard normal distribution, and $F^{-1}_{GOE(d)}$ is the inverse CDF of the largest eigenvalue of the Gausian Orthogonal Ensemble (GOE) matrix, whose distribution is calculated exactly by \citep{chiani2014distribution}. Algorithm ~\ref{alg:privacy_report} specializes to GLMs for clarity of presentation, but we could adapt it to any convex loss function by replacing the GLM-specific bounds with the more general ones.

We implicitly assume that the data analyst has already decided the privacy budgets $\epsilon_2$ and $\epsilon_3$ for the data-dependent release of the gradient (third term of $\epsilon_1(\cdot)$) and of the Hessian (first term of $\epsilon_1(\cdot)$). Inputs $\sigma_2$ and $\sigma_3$ are then calibrated to achieve $(\epsilon_2, \rho)$-DP and $(\epsilon_3, \rho)$-DP, respectively.


        \begin{algorithm}[H]
        \caption{Privacy report for \texttt{Obj-Pert} on GLMs}
        \label{alg:privacy_report}
        \begin{algorithmic}
            \STATE {\bfseries Input:} $\hat{\theta}^p \in \mathbb{R}^d$ from \texttt{Obj-Pert}, noise parameter $\sigma,\sigma_2, \sigma_3$; regularization parameter $\lambda$; Hessian $H := \sum_i\nabla^2\ell(\hat{\theta}^p; z_i) + \lambda I_d$, Boolean B $\in$ [\texttt{DATA-INDEP}, \texttt{DATA-DEP}], failure probability $\rho$
            \STATE {\bfseries Require: $\lambda \geq 2 \sigma_3 F_{\lambda_1(\mathrm{GOE}(d))}^{-1}(1-\rho/2)$}
            \STATE {\bfseries Output:} Reporting function $\tilde{\epsilon}: (x,y),\delta \rightarrow \R_+^3$
            \IF {B = \texttt{DATA-INDEP}}
            \STATE Set $\epsilon_2(\cdot):= 0, \epsilon_3(\cdot) := 0$.
            \vspace{2pt}
            \STATE Set $\overline{g^P}(z) := \sigma ||f'(\cdot)x||_2 F_{\cN(0,1)}^{-1}(1-\nicefrac{\rho}{2})$ and set $\overline{\mu^p}(x) := \frac{\|x\|^2}{\lambda}$. 
            \ELSIF {B = \texttt{DATA-DEP}}
            \STATE Privately release $\hat{g}^p$ by Algorithm ~\ref{alg:Alg_gp} with parameter $\sigma_2$.
            \STATE Set $\epsilon_2(\cdot)$ according to Theorem $\ref{thm:gp_pdp}$.
             \STATE Set {\small $\overline{g^P}(z) := \min\left\{f'(\cdot) [\hat{g}^P(z)]^Tx + \sigma_2 ||f'(\cdot)x||_2 F_{\cN(0,1)}^{-1}(1-\nicefrac{\rho}{2}),\; \sigma ||f'(\cdot)x||_2 F_{\cN(0,1)}^{-1}(1-\nicefrac{\rho}{2})\right\}.$}
            \STATE Privately release $\hat{H}^p$ by a variant of "Analyze Gauss"\footnotemark with parameter $\sigma_3$.
            \STATE Set $\epsilon_3(\cdot)$  according to Statement 2 of Theorem~\ref{thm:master_thm_reporting}.
            \STATE Set $\overline{\mu^p}(x) = \frac{3}{2}x^T [\hat{H}^p]^{-1} x$.
            \vspace{2pt}
            \ENDIF
            \vspace{2pt}
            \STATE Set $\overline{\epsilon_1^p}(z):= \big| -\log \big(1 -  f''(\cdot)\overline{\mu^p}(x) \big)  \big| + \frac{||f'(\cdot)x||_2^2}{2\sigma^2} +  \frac{\big|\overline{g^P}(z)\big|}{\sigma^2}.$
            \STATE Output the function $\tilde{\epsilon}(z) := \big(\overline{\epsilon_1^p}(z), \epsilon_2(z), \epsilon_3(z)\big)$.
            \end{algorithmic}
        \end{algorithm}
        \footnotetext{Instead of adding ``analyze-gauss'' noise, we sample from the Gaussian Orthogonal Ensemble (GOE) distribution to obtain a random matrix (Appendix~\ref{app:analyze_gauss}). Under this model we show that $\tau$ is on the order of $O(\sqrt{d}(1 + \log(C/\rho)^{3/2}))$.}

    Note that the pDP functions $\epsilon_2(\cdot)$ and $\epsilon_3(\cdot)$ -- which we use to report the additional pDP losses of releasing the private estimates of the gradient and the Hessian -- do not depend on the dataset, and thus are not required to be separately released. The privately released pDP functions depend on $\hat{\theta}^P$; to reduce clutter, we omit this parameter in our presentation of Algorithm ~\ref{alg:privacy_report}.

            \begin{theorem}\label{thm:master_thm_reporting}
    There is a universal constant $C$ such that if $\lambda > C\sigma_2\sqrt{d}(1+(\log(1/\rho))^{2/3})$, then Algorithm~\ref{alg:privacy_report} satisfies the following properties
    \begin{enumerate}
        \item $(\frac{\xi^2}{2\sigma_2^2}+\frac{\beta^2}{4\sigma_3^2} + \sqrt{\frac{\xi^2}{\sigma_2^2}+\frac{\beta^2}{2\sigma_3^3}}\sqrt{2\log(1/\delta)},\delta)$-DP
        \item $ ( \frac{f'(\hat{\theta}^p; z)^2\|x\|^2}{2\sigma_2^2} + \frac{f''(\hat{\theta}^p; z)^2 \|x\|^4}{4\sigma_3^2} + \sqrt{\frac{f'(\hat{\theta}^p; z)^2\|x\|^2}{\sigma_2^2} + \frac{f''(\hat{\theta}^p; z)^2\|x\|^4}{2\sigma_3^2}} \sqrt{2\log(1/\delta)},\delta)$-pDP for all $x\in\cX$ and $0 \leq \delta<1$.

        \item   For a fixed input $z$ and $D$, and all $\rho>0$, the privately released privacy report $\tilde{\epsilon}(\cdot)$ satisfies that  
        $\epsilon_1(\hat{\theta}^p, D, D_{\pm z}) \leq \overline{\epsilon^p_1}(z)
        \leq 12\epsilon_1(\hat{\theta}^p, D, D_{\pm z}) + \frac{|f'(\cdot)|\|x\|}{\sigma_2}\sqrt{2 \log(2/\rho)}$ with probability $1-3\rho$ where $\epsilon_1(\cdot)$ is the expression from Theorem~\ref{thm:pdp_alg1_general}.
    \end{enumerate}
    \end{theorem}
    
    
    
    
    \noindent\textbf{Accurate approximation with low privacy cost.} 
    This theorem shows that if we 
    use a slightly larger $\lambda$ in ObjPert then we get an upper bound of the pDP for each individual $z$ up to a multiplicative and an additive factor.  The multiplicative factor is coming from a multiplicative approximation of $-\log \big(1 \pm f''(\cdot)\mu(x) \big)$ and the additive error is due to the additional noise added for releasing the third term $\frac{1}{\sigma^2} \nabla J(\hat{\theta}^P; D)^T\nabla \ell(\hat{\theta}^P; z)$.
    The additional DP and pDP losses for releasing $H$ and $g$ are comparable to the DP and pDP losses in Objective Perturbation itself if $\sigma_2\asymp \sigma_3 \asymp \sigma$.
     
    Moreover, while using a large $\lambda$ may appear to introduce additional bias, the required choice of $\lambda \asymp \sqrt{d}\sigma$ is actually exactly the choice to obtain the minimax rate in general convex private ERM \citep{bassily2014private} (Figure~\ref{fig:lambda_utility} demonstrates the impact of increasing $\lambda$). 
    
    
    \noindent\textbf{Joint DP interpretation.} Finally, we can also interpret our results from a joint-DP perspective \citep{kearns2014mechanism}.  Given any realized output $\hat{\theta}^p\in \R^d$, the tuple of $\{\tilde{\epsilon}(z_1,\hat{\theta}^p),...,\tilde{\epsilon}(z_1,\hat{\theta}^p)\}$ satisfies joint DP with the same $\epsilon$ parameter as in Theorem~\ref{thm:master_thm_reporting}. This follows from the billboard lemma \citep{hsu2016private}.

\section{Experiments}
\label{experiments}

Here we evaluate our methods to release the pDP losses using logistic regression as a case study. In Section ~\ref{reg_utility}, we demonstrate that the stronger regularization required by Algorithm ~\ref{alg:privacy_report} does not affect the utility of the model. In Section ~\ref{dep_indep_compare} we show that by carefully allocating the privacy budget of the data-dependent release, we can achieve a more accurate estimate of the \emph{ex-post} pDP losses of Algorithm ~\ref{alg:Alg1} compared to the data-independent release, with reasonable overhead (same overall DP budget and only a slight uptick in the overall pDP losses).

Experiments with linear regression, with additional datasets and with alternative privacy budget allocation schemes are included in the supplementary materials.
\subsection{Stronger regularization does not worsen model utility}
\label{reg_utility}

In this experiment we use a synthetic dataset generated by sampling $x_i, \theta \sim \mathcal{N}(0, I_d)$ and normalizing each $x_i \in X$ so that $||x_i||_2 = 1$. Then we rescale $Y = X \theta$ to ensure $y_i \in [0, 1]$ for each $y_i \in Y$. \newline\newline
\begin{minipage}{.4\textwidth}
Algorithm ~\ref{alg:privacy_report} requires a larger $\lambda$ than suggested by Theorem $\ref{thm:alg1}$ in order to achieve a uniform multiplicative approximation of $\mu(\cdot)$.  
We investigate the effect of stronger regularization on the utility of a private logistic regression model applied to a synthetic dataset ($n = 1000, d = 50$), for several settings of $\epsilon_1$. Since each value of $\epsilon_1$ demands a different minimum value of $\lambda$ in order to achieve $(\epsilon_1, \delta)$-differential privacy, we compare via "$\lambda$-inflation": a measure of how many times larger we set $\lambda$ than its minimum value required to achieve the worst-case DP bound of objective perturbation.
\end{minipage}
\hspace{.2cm}\begin{minipage}{.6\textwidth}
\begin{figure}[H]
  \centering
    \includegraphics[width=\textwidth]{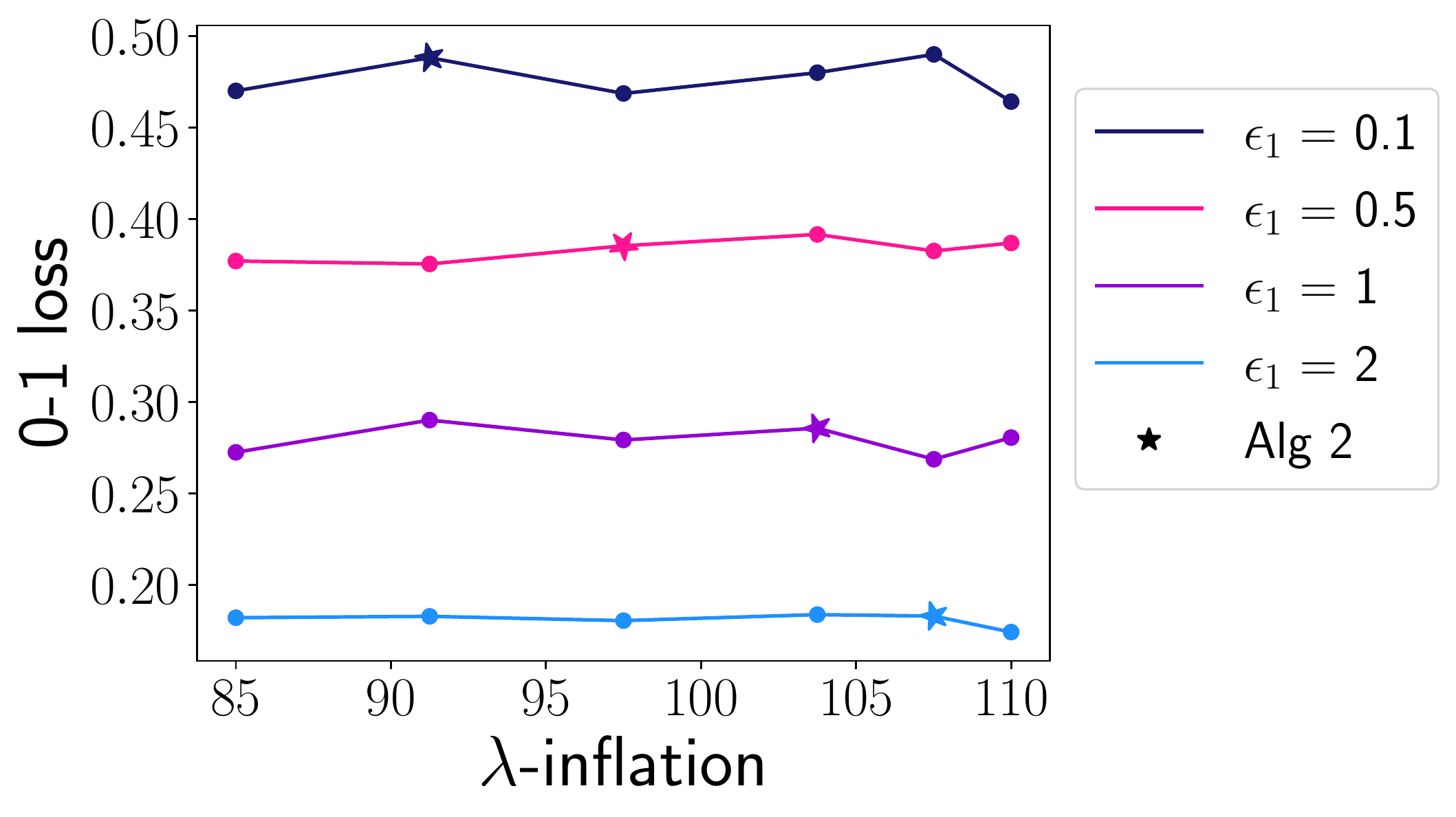}
  \caption{Utility of Obj-Pert with larger $\lambda$.}
  \label{fig:lambda_utility}
\end{figure}
\end{minipage}

  E.g., for logistic regression the objective perturbation mechanism requires $\lambda \geq \frac{1}{2\epsilon}$, and so in Figure ~\ref{fig:lambda_utility} a $\lambda$-inflation value of 10 means that we set $\lambda = \frac{5}{\epsilon}$. For each $\lambda$-inflate value $c$ , we run Algorithm \ref{alg:Alg1} with $\lambda = c \lambda_{\texttt{Obj-Pert}}$.  In particular, the star symbol marks the level of $\lambda$-inflation enforced by Algorithm ~\ref{alg:privacy_report}. The experimental results summarized in Figure ~\ref{fig:lambda_utility} show that the performance of the private logistic regression model (as measured by the 0-1 loss) remains roughly constant across varying scales of $\lambda$.

\subsection{Comparison of data-independent and data-dependent bounds}
\label{dep_indep_compare}

The following experiments feature the credit card default dataset ($n = 30000, d = 21$) \citep{yeh2009comparisons} from the UCI Machine Learning Repository. We privately train a binary classifier to predict whether or not a credit card client $z$ defaults on her payment (Algorithm \ref{alg:Alg1}), and calculate the true pDP loss $\epsilon_1(\cdot)$ as well as the data-independent and -dependent estimates $\overline{\epsilon_1^P}(\cdot)$ for each $z$ in the training set (Algorithm \ref{alg:privacy_report}).

The failure probabilities for both Algorithms ~\ref{alg:Alg1} and ~\ref{alg:privacy_report} are set as $\delta = \rho = 10^{-6}$. Our choices of $\sigma$ and $\lambda$ depend on $\epsilon_1$ and follows the requirements stated in Theorem \ref{thm:alg1} to achieve DP. We don't use any additional regularization, i.e. $r(\theta) = 0$. For the data-dependent release, the noise parameters $\sigma_2, \sigma_3$ are each calibrated according to the analytic Gaussian mechanism of \citep{balle2018improving}.

\begin{figure}[H]
  \centering
    \includegraphics[width=\textwidth]{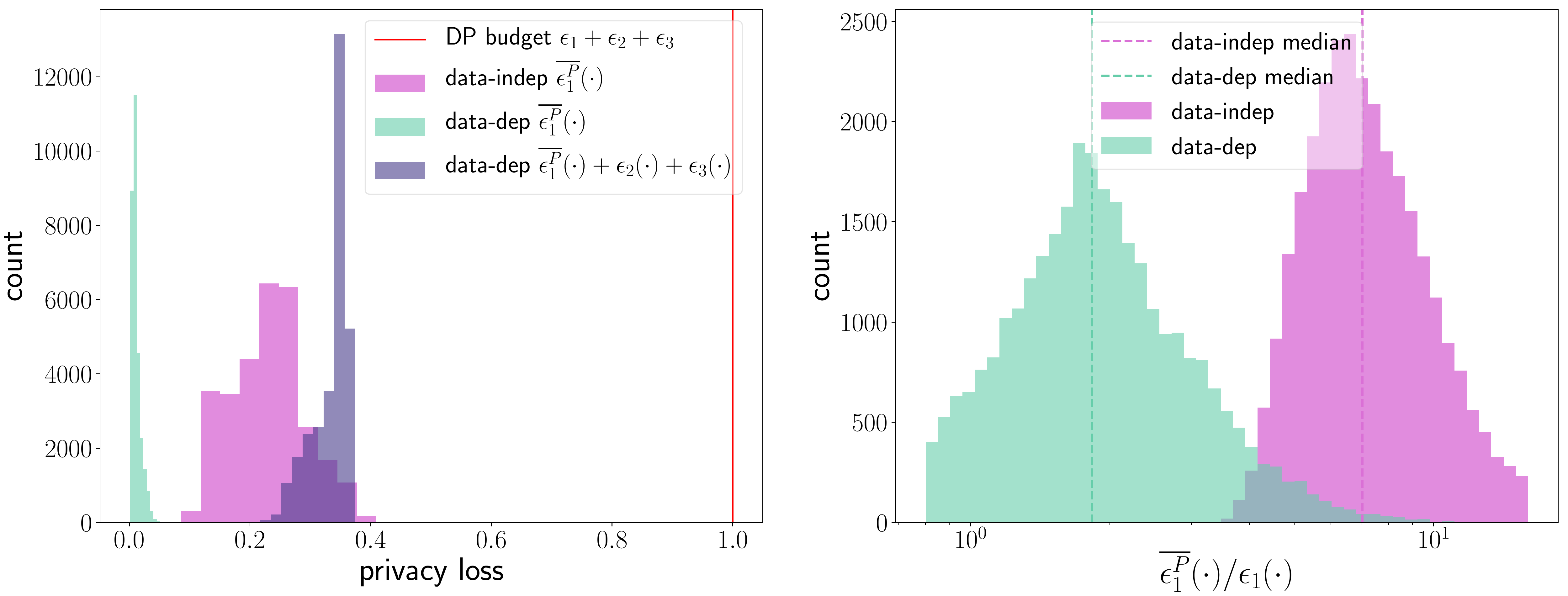}
\subfloat[Distribution of privately released \emph{ex-post} pDP losses. \label{fig:ep1_align_raw}]{\hspace{.39\linewidth}}\hspace{1.5cm}
\subfloat[ Distribution of ratios between the privately released \emph{ex-post} pDP losses $\overline{\epsilon_1^P}(\cdot)$ and their true values $\epsilon_1(\cdot)$. \label{fig:comp_ratio}]{\hspace{.35\linewidth}}
  \caption{True and privately released pDP losses when the total privacy budget is $\epsilon = 1$. For the data-independent release we use the entire privacy budget on releasing $\hat{\theta}^P$ ($\epsilon_1 = 1$). For the data-dependent release we reserve some of the privacy budget for releasing $\overline{\mu^P}(\cdot)$ and $\overline{g^P}(\cdot)$ ($\epsilon_1 = 0.2, \epsilon_2 = 0.7, \epsilon_3 = 0.1$).}
  \label{fig:ep1_aligned}
\end{figure}

Using $\epsilon = 1$ as a DP budget, we investigate how to allocate the privacy budget among the components of the data-dependent release ($\epsilon = \epsilon_1 + \epsilon_2 + \epsilon_3$) to achieve a favorable comparison with the data-independent release which requires no additional privacy cost ($\epsilon = \epsilon_1$). The configuration described in Figure ~\ref{fig:ep1_aligned}, which skews the data-dependent privacy budget toward more accurately releasing $\overline{\epsilon_1^P}(\cdot)$, was empirically chosen as an example where the sum $\overline{\epsilon_1^P}(\cdot) + \epsilon_2(\cdot) + \epsilon_3(\cdot)$ of privately released pDP losses of the data-dependent approach  are comparable to the privately released $\emph{ex-post}$ pDP loss $\overline{\epsilon_1^P}(\cdot)$ of the data-independent approach. Note that $\epsilon_2(\cdot)$ and $\epsilon_3(\cdot)$ aren't $\emph{ex-post}$ in the traditional sense; however, we feel comfortable summing $\overline{\epsilon_1^P}(\cdot) + \epsilon_2(\cdot) + \epsilon_3(\cdot)$ since all three terms are a function of $\hat{\theta}^P$ and individual $z$'s data. Note also that since the total budget $\epsilon$ is the same for both the data-independent and -dependent releases, $\epsilon_1$ differs between them. Therefore Figure ~\ref{fig:comp_ratio} compares the accuracy of both approaches using the ratio between $\overline{\epsilon_1^P}(\cdot)$ and $\epsilon_1(\cdot)$ rather than their raw values.

When including the additional privacy budget incurred by the data-dependent approach, the data-dependent approach loses its competitive edge over the data-independent approach. Note that setting $\epsilon_2 = \epsilon_3 = 0$ would reduce the data-dependent approach to the data-independent one. The real advantage of the data-dependent approach can be best seen by allotting only a small portion of the overall privacy budget to Algorithm ~\ref{alg:Alg1}; then we can release $\hat{\theta}^P$ and $\overline{\epsilon_1^P}(\cdot)$ with reasonable overhead while achieving tighter and more accurate upper bounds for $\overline{\mu^P}(\cdot)$ and $\overline{g^P}(\cdot)$. By suffering a small additional \emph{ex-post} pDP loss (Figure ~\ref{fig:ep1_align_raw}), we can release the \emph{ex-post} pDP losses of Algorithm ~\ref{alg:Alg1} much more accurately (Figure ~\ref{fig:comp_ratio}). The downside to this is that reducing $\epsilon_1$ reduces the accuracy of the output $\hat{\theta}^P$. Deciding how to allocate the privacy budget between $\epsilon_1, \epsilon_2$ and $\epsilon_3$ thus requires weighing the importance of an accurate $\hat{\theta}^P$ against the importance of an accurate $\overline{\epsilon_1^P}(\cdot)$.

\section{Conclusion}
\label{conclusions}

We precisely calculate the privacy loss that an individual $z$ suffers \emph{after} the objective perturbation algorithm is run on a specific dataset. The \emph{ex-post} pDP loss function in DP-ERM can be accurately released to all individuals with little or
no additional privacy cost. In particular, we present a \emph{data-independent} bound which empirically provides a reasonably accurate estimate of the \emph{ex-post} pDP loss while requiring no further privatization step. Reserving some of the privacy budget allows us to alternatively release a tighter \emph{data-dependent} bound.

An important next step is to extend the per-instance DP analysis to the setting of constrained optimization. There are many promising future directions including using publishable pDP losses for designing more data-adaptive DP algorithms.

\subsection*{Acknowledgments}
The work was partially supported by NSF CAREER Award \# 2048091,  Google Research Scholar Award and a gift from Evidation Health. 

\bibliography{p4}
\bibliographystyle{p4_style}

\newpage
\appendix

\addcontentsline{toc}{section}{Appendix} 
\part{Appendix} 
\parttoc 

\newpage
\section{DP Variants}
\label{app:dp_variants}
\newcommand\Tstrut{\rule{0pt}{1.3\normalbaselineskip}}
\newcommand\bigTstrut{\rule{0pt}{1.7\normalbaselineskip}}
\newcommand\reallybigTstrut{\rule{0pt}{2.4\normalbaselineskip}}

\newcommand\Bstrut{\rule[-1.3\normalbaselineskip]{0pt}{0pt}}   \newcommand\bigBstrut{\rule[-2\normalbaselineskip]{0pt}{0pt}}

Algorithm design is a typical use case for differential privacy: given a privacy budget of $\epsilon$, the data curator would like to add noise calibrated to meet the privacy demands. Our work concerns the converse problem of how to calculate and report the \emph{incurred} privacy loss to an individual after a randomized algorithm is run on a fixed dataset. The table below summarizes the relevant variations of the DP definition which characterize the privacy loss with varying degrees of granularity. 

Let $P, Q$ be distributions over $\Omega$, taking $p(\omega)$ and $q(\omega)$ to be the probability density/mass function of each at $\omega$. Then the probability metrics used in the table are defined as follows:
\begin{itemize}
    \item $D_{\infty}(P||Q) = \sup\limits_{S\subset \Omega} \left(\log \dfrac{P(S)}{Q(S)}\right)$ \:\:(max divergence)
    \item $D_{\infty}^{\delta}(P||Q)  = \sup\limits_{S\subset \Omega: P(\omega) \geq \delta} \left(\log \dfrac{P(S) - \delta}{Q(S)}\right)$ \:\:($\delta$-approximate max divergence),
    \item $D_{\alpha}(P||Q) = \dfrac{1}{\alpha - 1} \log 
\E_{\omega \sim Q} \bigg[\bigg( \dfrac{p(\omega)}{q(\omega)} \bigg)^{\alpha}\bigg]$ \:\:(R\'enyi divergence).
\end{itemize}

\begin{center}
\begin{tabular}{ |c|c|c| } 
 \hline 
  \Tstrut \textbf{Pure DP} & $\sup\limits_D \sup\limits_{{z, D': D' \simeq_z D}} D_{\infty}\big(\mathcal{A}(D)|| \mathcal{A}(D')\big) \leq \epsilon$ \Bstrut   \\ \hline
  \Tstrut \textbf{Approximate DP} & $\sup\limits_D \sup\limits_{z, D': D' \simeq_z D} D_{\infty}^{\delta}\big(\mathcal{A}(D)|| \mathcal{A}(D')\big) \leq \epsilon$\Bstrut  \\ \hline
 \Tstrut \textbf{R\'enyi DP} & $\sup\limits_D \sup\limits_{z, D': D' \simeq_z D} D_{\alpha}\big(\mathcal{A}(D)|| \mathcal{A}(D')\big) \leq \epsilon$\Bstrut  \\ \hline
 \Tstrut \textbf{Data-dependent DP} & $ \sup\limits_{z, D': D' \simeq_z D} D_{\alpha}\big(\mathcal{A}(D)|| \mathcal{A}(D')\big) \leq \epsilon(D)$\Bstrut \\ \hline
 \bigTstrut \textbf{Personalized DP} & $ \sup\limits_{D, D': D' \simeq_z D} \max \Big(D_{\infty}^{\delta}\big(\mathcal{A}(D)|| \mathcal{A}(D')\big), D_{\infty}^{\delta}\big(\mathcal{A}(D')|| \mathcal{A}(D)\big)\Big) \leq \epsilon(z)$\Bstrut  \\ \hline
 \bigTstrut \textbf{Per-instance DP} & $ \max \Big(D_{\infty}^{\delta}\big(\mathcal{A}(D)|| \mathcal{A}(D')\big), D_{\infty}^{\delta}\big(\mathcal{A}(D')|| \mathcal{A}(D)\big)\Big) \leq \epsilon(D, z)$\Bstrut  \\ \hline
  \reallybigTstrut \textbf{\textit{Ex-post} per-instance DP} & $\left| \log \dfrac{\text{Pr}\big[\mathcal{A}(D) = o \big]}{\text{Pr}\big[\mathcal{A}(D') = o \big]} \right| \leq \epsilon(o, D, D')$ \: where $D' \simeq_z D$\bigBstrut \\ \hline
\end{tabular}
\end{center}

\newpage
\section{Additional Experiments}
\label{app:additional_experiments}

\subsection{Varying dimension and dataset size}

Our first experiment uses a synthetic dataset for logistic regression as described in the experiments section of the main paper. Figure ~\ref{fig:ep1_vary_n} illustrates how the worst-case pDP loss over all individuals in the dataset -- i.e., $\max_{z \in D} \epsilon_1(\hat{\theta}^P, D, D_{\pm z})$ -- changes as a function of the dataset size (number of individuals in the dataset) $n$, compared to the worst-case pDP bounds given by the data-independent and data-dependent approaches. We fix $d = 50$ and vary $n$ from $n = 100$ to $n = 10000$.

Figure ~\ref{fig:ep1_vary_n} illustrates how the worst-case pDP loss and bounds change as a function of the data dimension $d$. We fix $n = 1000$ and vary $d$ from $d = 1$ to $d = 60$. Figures ~\ref{fig:ep1_vary_n} and ~\ref{fig:ep1_vary_d} demonstrate that for GLMs, the strength of our \emph{ex-post} pDP bounds $\epsilon_1^P(\cdot)$ does not depend on the size of the dataset or the dimensionality of the data.

\begin{minipage}{.5\textwidth}
\begin{figure}[H]
  \centering
    \includegraphics[width=\textwidth]{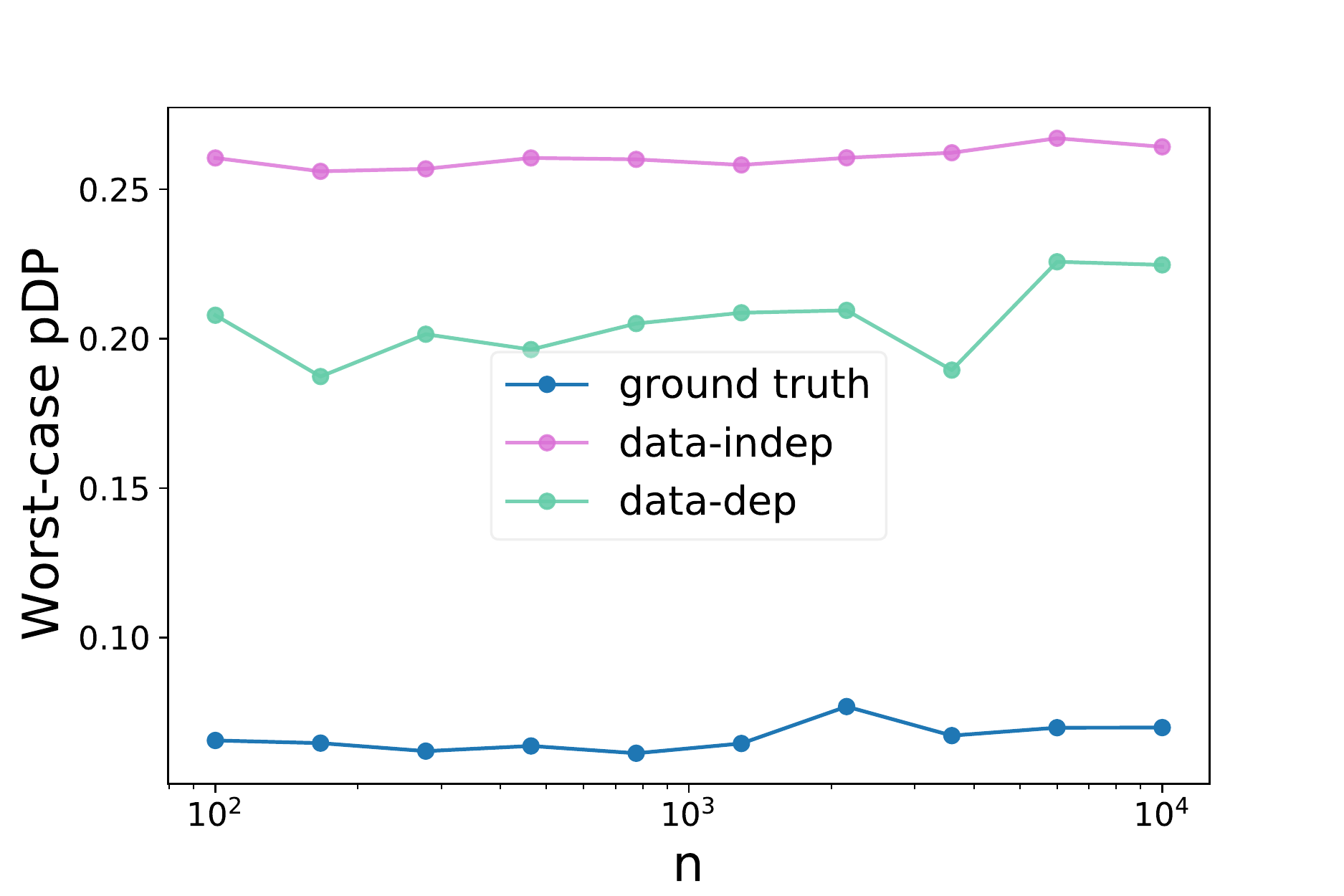}
  \caption{Worst-case pDP while varying $n$.}
  \label{fig:ep1_vary_n}
\end{figure}
\end{minipage}
\begin{minipage}{.5\textwidth}
\begin{figure}[H]
  \centering
    \includegraphics[width=\textwidth]{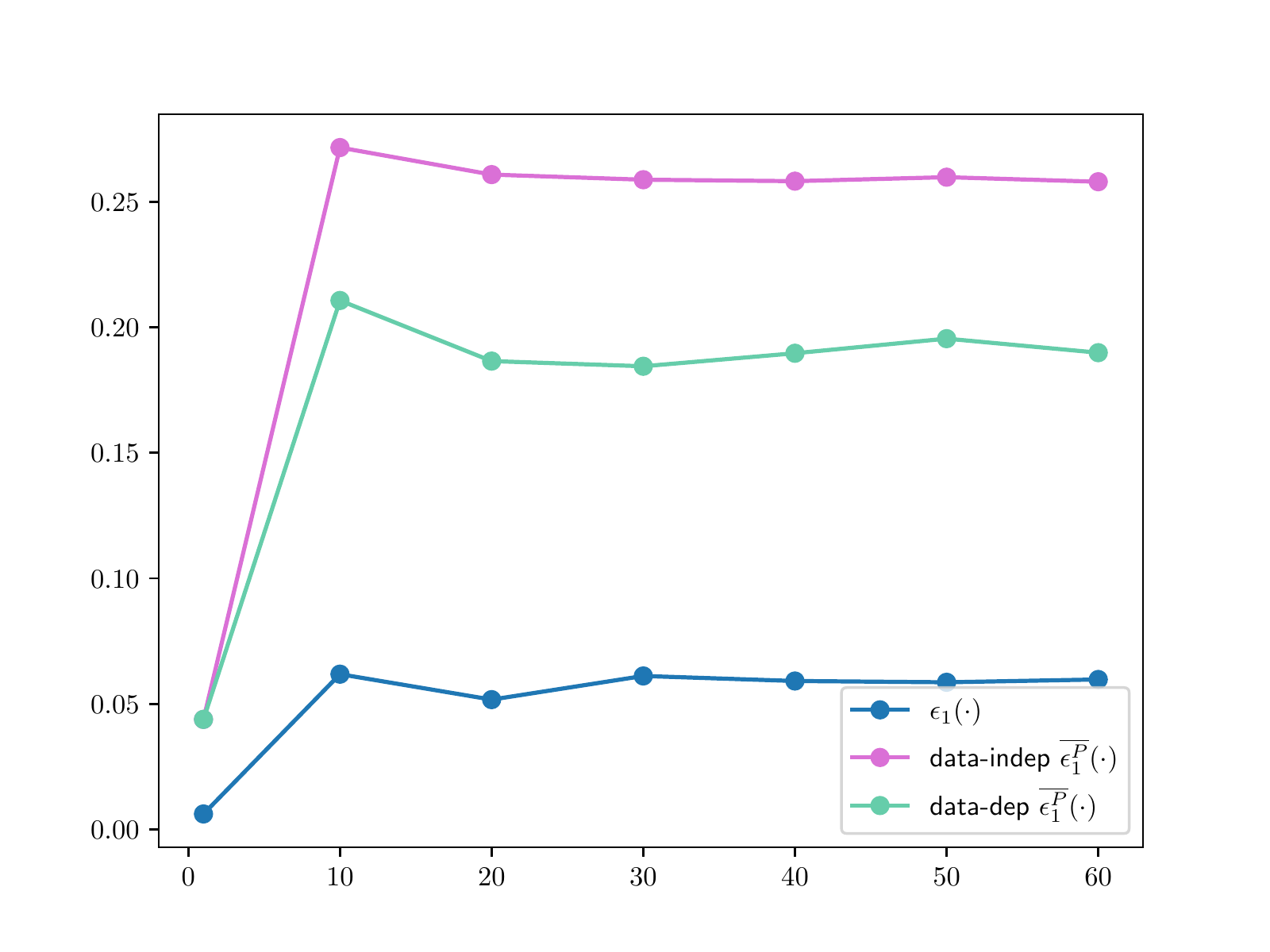}
  \caption{Worst-case pDP while varying $d$.}
  \label{fig:ep1_vary_d}
\end{figure}
\end{minipage}

\subsection{Privacy budget allocation}

Here we investigate how to distribute the privacy budget between the components of Algorithm ~\ref{alg:Alg1} and Algorithm ~\ref{alg:privacy_report}, with the same experimental setup as in Section ~\ref{dep_indep_compare}. As before, we use the UCI credit default dataset. Our experiments show that a careful allocation of the privacy budget is essential to reaping the benefits of the data-dependent approach to releasing the \emph{ex-post} pDP losses. 

The plots in Figure ~\ref{fig:ep1_aligned} are ordered by increasing $\epsilon_1^{DEP}$.  $\epsilon_1^{INDEP} = 1$ is fixed, as are (implicitly) $\epsilon_2^{INDEP} = \epsilon_3^{INDEP}=  0$. We see that as $\epsilon_1^{DEP}$ approaches the total privacy budget of $\epsilon_1^{INDEP} = 1$, leaving less budget for $\epsilon_2^{DEP}$ and $\epsilon_3^{DEP}$, the data-dependent release is little better than the data-independent release -- worse, even, because we've expended additional privacy cost without significantly boosting the accuracy of the release.

Deciding between the data-independent or data-dependent approach is a delicate choice which depends on the particular problem setting. However, based on our theoretical and experimental results we can offer some loose guidelines:
\begin{itemize}
    \item For non-GLMs, the data-independent bound has a dimension dependence. Therefore in the high-dimensional case, we recommend the data-dependent approach for generic convex loss functions and the data-independent approach for GLMs.
    \item For GLMs, the data-independent approach gives tight bounds without any overhead. The only reason to use the data-dependent approach for GLMs would be to gain an even more accurate estimate of the \emph{ex-post} pDP losses, in which case it would be necessary to either suffer an additional privacy cost, or maintain the privacy cost by suffering a less accurate estimate of $\hat{\theta}^P$.
\end{itemize}

\begin{figure}[H]
  \centering
    \subfloat{\includegraphics[width=\textwidth]{experiments/align_combined.pdf}} \\
    \subfloat{\includegraphics[width=\textwidth]{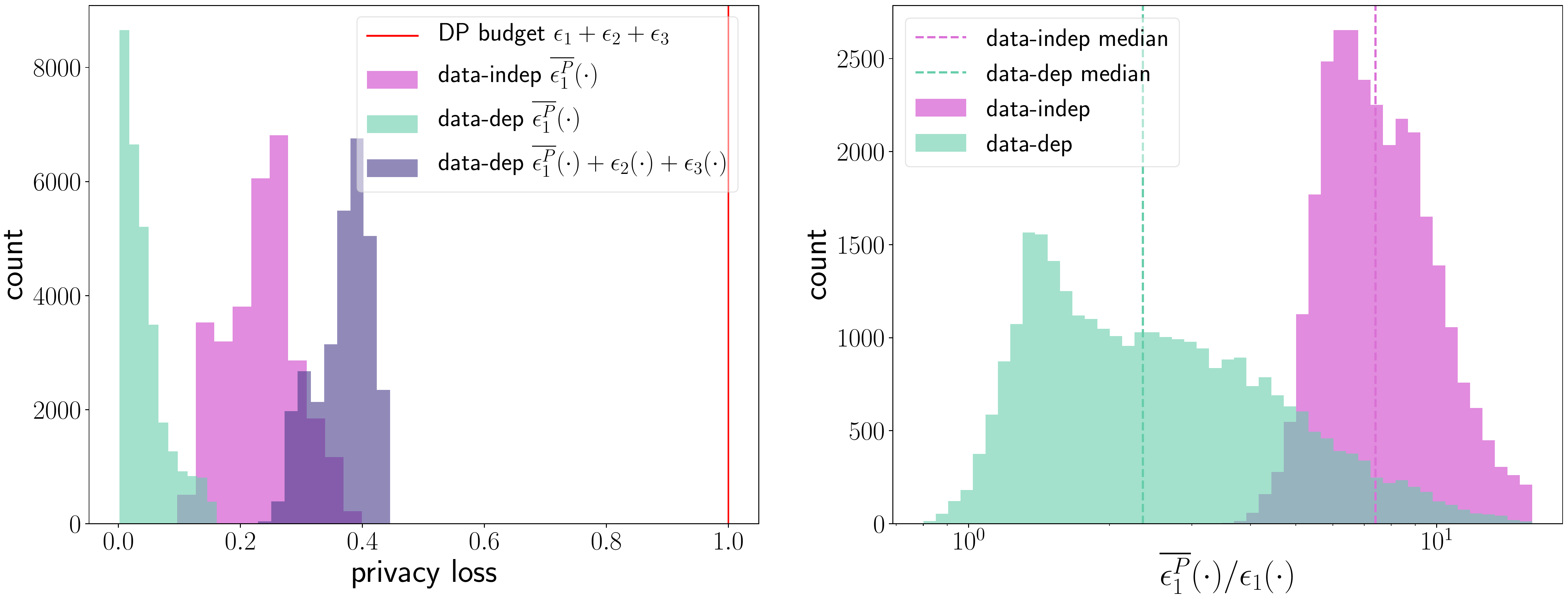}}\\
    \subfloat{\includegraphics[width=\textwidth]{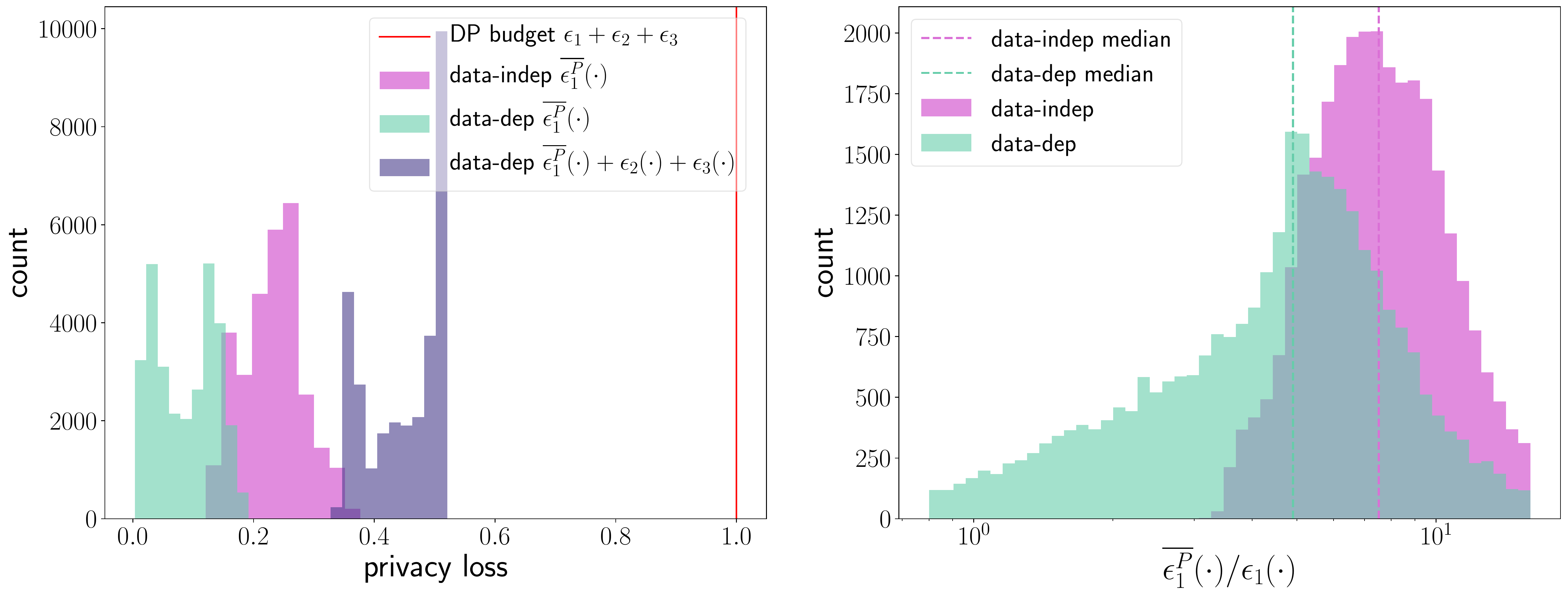}} \\
    \subfloat{\includegraphics[width=\textwidth]{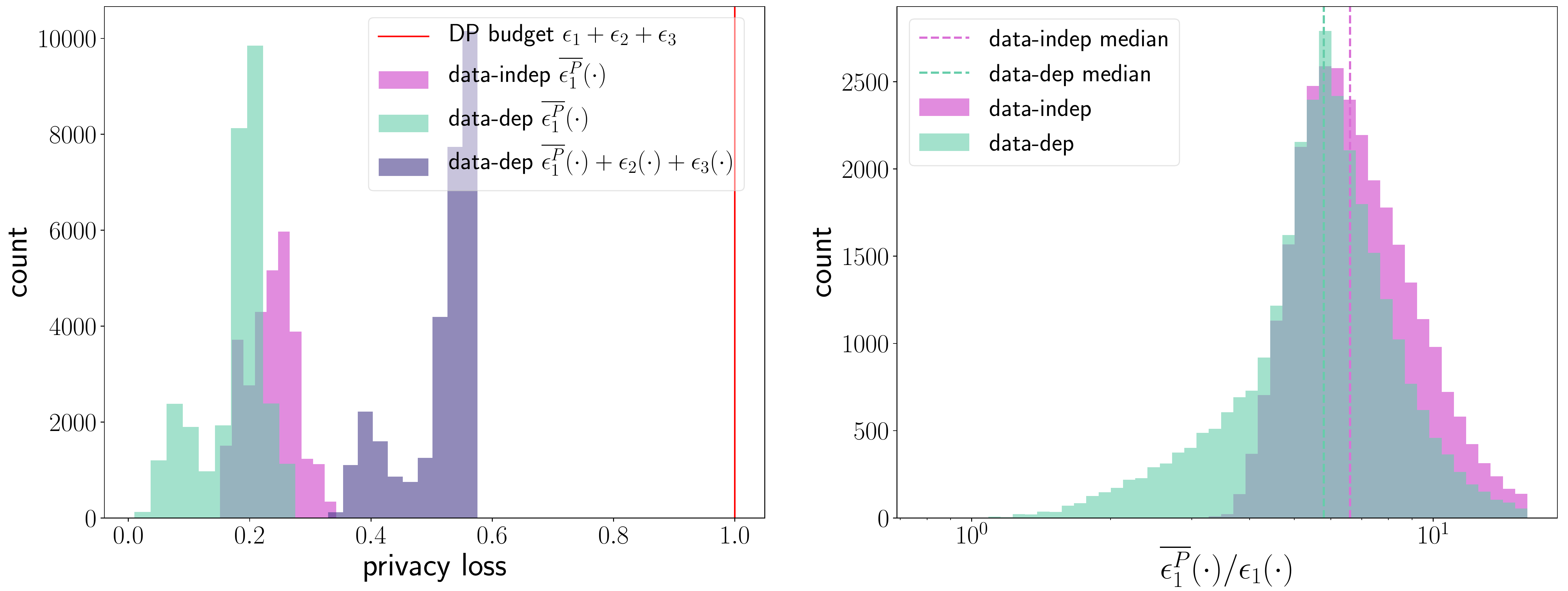}}
    \caption{Data-independent release uses a privacy budget $\epsilon_1 = 1$ for each plot. From top to bottom, the budgets for the data-dependent release are $\epsilon_1 = 0.2, \epsilon_2 = 0.7, \epsilon_3 = 0.1; \epsilon_1 = 0.4, \epsilon_2 = 0.5, \epsilon_3 = 0.1; \epsilon_1 = 0.5, \epsilon_2 = 0.25, \epsilon_3 = 0.25;$ and $\epsilon_1 = 0.8, \epsilon_2 = 0.1, \epsilon_3 = 0.1$.}

\label{fig:ep1_aligned}
\end{figure}

\subsection{Comparison of pDP losses and private upper bounds}

\begin{figure}[H]
  \centering
    \includegraphics[width=\textwidth]{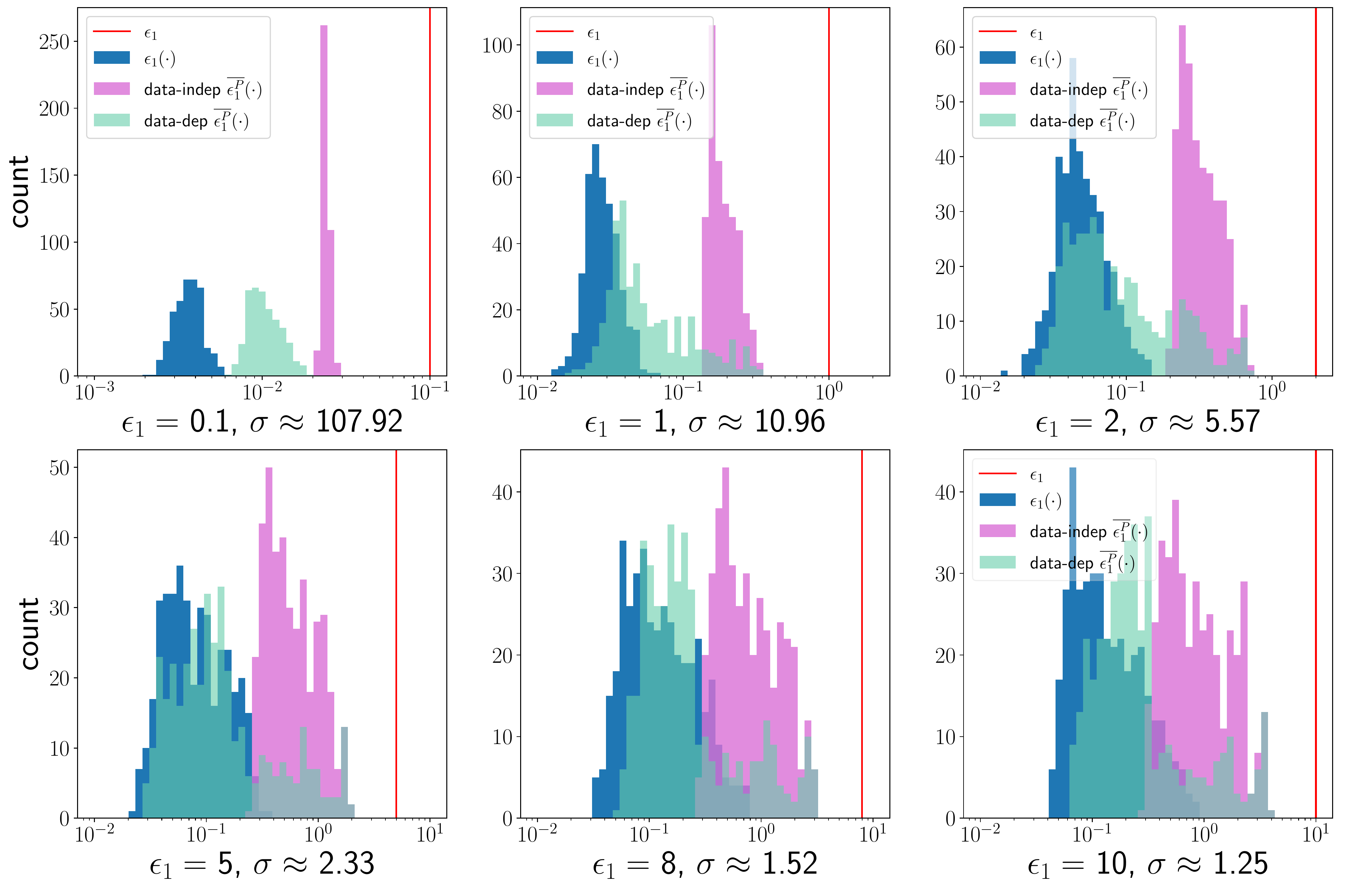}
  \caption{pDP losses $\epsilon_1(\cdot)$ and upper bounds $\overline{\epsilon_1^P}(\cdot)$ for private logistic regression applied to the UCI kidney dataset. DP budget for releasing $\hat{\theta}^P$ is $\epsilon = 1$, marked in red.}
  \label{fig:ep1_kidney}
\end{figure}

\begin{figure}[H]
  \centering
    \includegraphics[width=\textwidth]{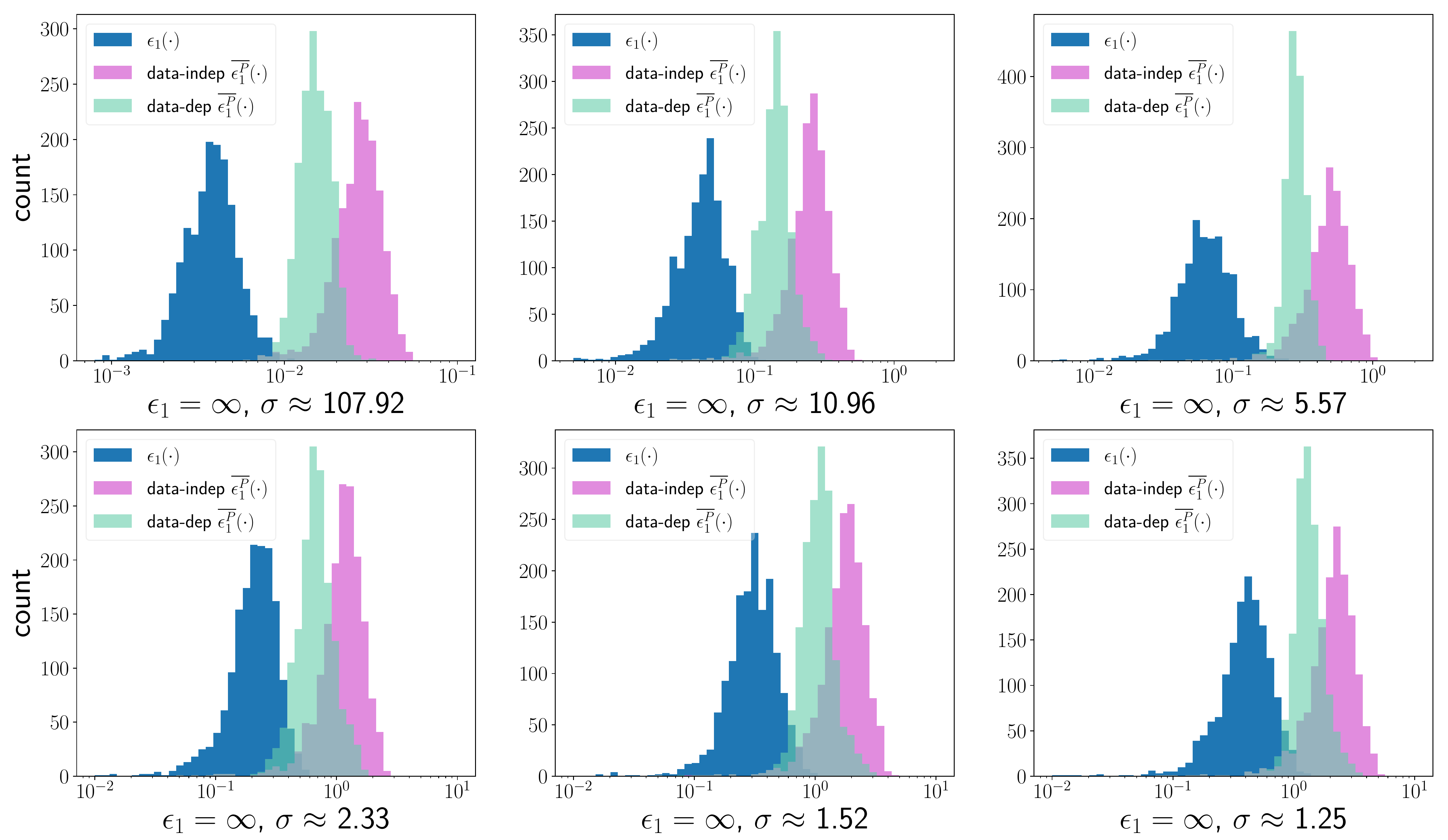}
  \caption{pDP losses $\epsilon_1(\cdot)$ and upper bounds $\overline{\epsilon_1^P}(\cdot)$ for private linear regression applied to the UCI wine quality dataset. Since we are dealing with an unbounded domain $\mathbb{R}^d$,  the algorithm does not satisfy worst-case DP for any $\epsilon < \infty$.}
  \label{fig:ep1_wine}
\end{figure}

We run both the data-independent and -dependent variations of Algorithm ~\ref{alg:Alg1} as described in the experimental setup. Note that in this experiment the additional DP budget for the data-dependent release is $\epsilon_2 = \epsilon_3 = 1$, i.e. the privacy budget for the data-dependent release is three times the DP budget for the data-independent release.Figures ~\ref{fig:ep1_kidney} and ~\ref{fig:ep1_wine} compare the pDP losses $\epsilon_1(\cdot)$ and private upper bounds $\overline{\epsilon_1^P}$ with $\epsilon_1$ (indicated by the vertical red line), the DP budget for Algorithm ~\ref{alg:Alg1}. Figure ~\ref{fig:ep1_kidney} shows results for private logistic regression on the UCI kidney dataset; Figure ~\ref{fig:ep1_wine} shows results for private linear regression on the UCI wine quality dataset \citep{Dua:2019}. Our experimental results indicate that for smaller $\epsilon_1 << 1$ (larger $\sigma$), the data-dependent approach provides a markedly tighter bound on $\epsilon_1()\cdot$.

\begin{figure}[H]
  \centering
    \includegraphics[width=\textwidth]{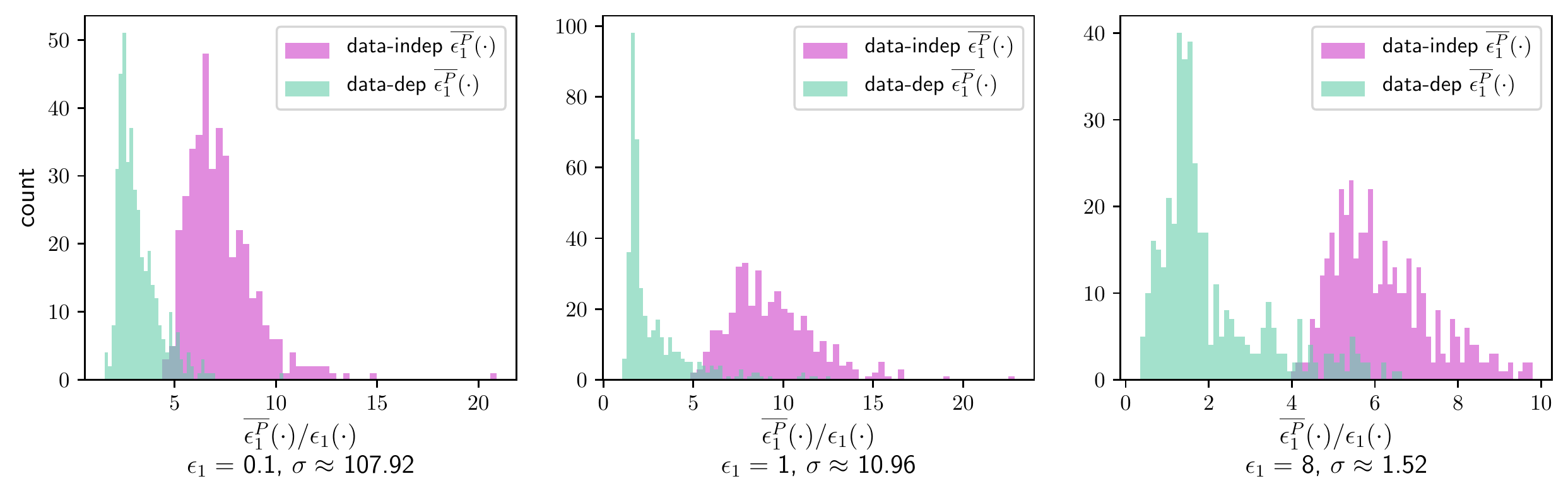}
  \caption{Ratio of private upper bound $\overline{\epsilon_1^P}(\cdot)$ to actual pDP loss $\epsilon_1(\cdot)$ for private logistic regression applied to the UCI kidney dataset.}
  \label{fig:ep1_kidney_ratio}
\end{figure}

\begin{figure}[H]
  \centering
    \includegraphics[width=\textwidth]{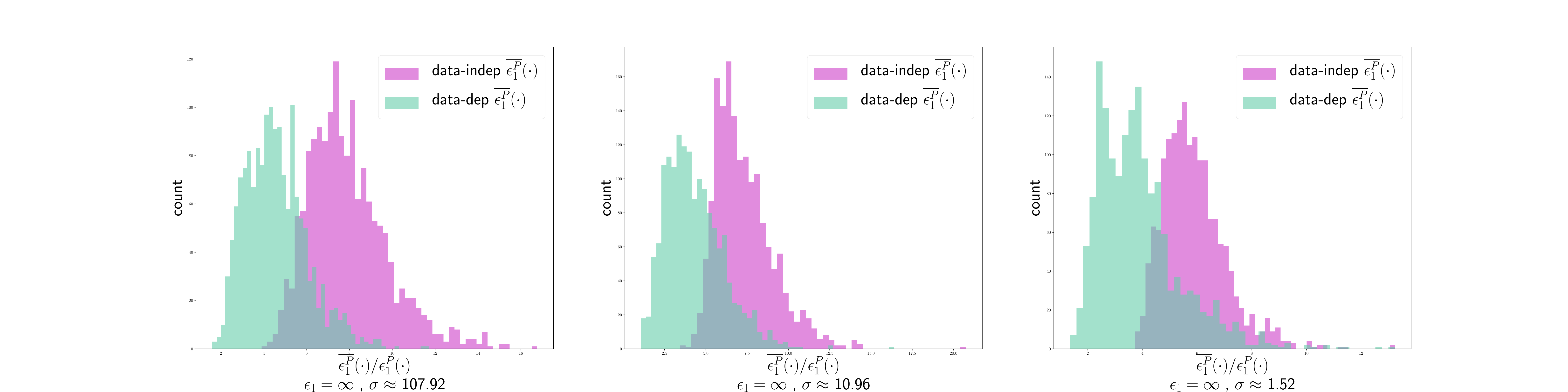}
  \caption{Ratio of private upper bound $\overline{\epsilon_1^P}(\cdot)$ to actual pDP loss $\epsilon_1(\cdot)$ for private linear regression applied to the UCI wine quality dataset.}
  \label{fig:ep1_wine_ratio}
\end{figure}

Figures ~\ref{fig:ep1_kidney_ratio} and ~\ref{fig:ep1_wine_ratio} plot the ratio of the private upper bound $\epsilon_1^P(\cdot)$ for both the data-independent and -dependent approaches to the true pDP loss $\epsilon_1(\cdot)$. This illustrates the relative accuracy of the pDP estimates $\epsilon_1^P(\cdot)$. For both logistic regression on the UCI kidney dataset (Figure ~\ref{fig:ep1_kidney_ratio}) and linear regression on the UCI wine quality dataset (Figure ~\ref{fig:ep1_wine_ratio}), the data-dependent approach provides a more accurate estimate of the pDP loss $\epsilon_1(\cdot)$, especially for logistic regression on the kidney dataset.


\newpage
\section{Even Stronger Privacy Report}
\label{app:adaptive_report}


\subsection{More Accurate Privacy Report by Adapting to the Data }
We now present a more adaptive version of Algorithm ~\ref{alg:privacy_report} that could be even more accurate depending on the intrinsic stability of the dataset itself. The key technical components include:
\begin{itemize}
    \item Adapting to a well-conditioned $H$ by releasing $\lambda_{\min}$.
    \item A ``regularized'' construction of $\hat{\mu}^p(\cdot)$ that provides valid upper bounds of $\mu(\cdot)$ for all choices of $\lambda > 0$.
\end{itemize}

Algorithm ~\ref{alg:adaptive_report} makes use of a subroutine to add noise to the smallest eigenvalue of $H$, presented below along with its privacy guarantees.

  \begin{algorithm}[H]
                        \caption{ Releasing the smallest eigenvalue of $H$ }                \label{alg:lambda_min}
        \begin{algorithmic}
            \STATE{{\bfseries Input:} Dataset $D$, noise parameter $\sigma_4$, $\lambda_{\min}$ denoting the smallest eigenvalue of $H$.}
            \STATE{{\bfseries Output:} $\hat{\lambda}_{\min}^P$.}
           \STATE{Output $\hat{\lambda}_{\min}^P = \lambda_{\min} + \cN(0,\sigma_4^2)$ .}
        \end{algorithmic}
    \end{algorithm}

\begin{theorem}\label{thm:pdp_of_lambda_min}
    Algorithm~\ref{alg:lambda_min} satisfies pDP with 
    $$
    \epsilon_4(\cdot) = \frac{f^{''}(\cdot)^2\|x\|^4}{2\sigma_4^2} +  \frac{f^{''}(\cdot)\|x\|^2\sqrt{2\log(1/\delta)}}{\sigma_4},
    $$
    and if $f^{''}(\cdot)\|x\|^2 \leq \beta$ for all $x$ then Algorithm~\ref{alg:lambda_min} also satisfies $(\epsilon,\delta)$-DP with $\epsilon = \frac{\beta^2}{2\sigma_4^2} +  \frac{\beta\sqrt{2\log(1/\delta)}}{\sigma_4}$.
\end{theorem}
\begin{proof}
Algorithm~\ref{alg:lambda_min} is a standard Gaussian mechanism. By Weyl's lemma, the smallest singular value satisfies a perturbation bound of $f''(\hat{\theta}^p; z)\|xx^T\|_2 = f''(\hat{\theta}^p; z)\|x\|^2$ from adding or removing one individual data point. The stated result follows from the theorem of the Gaussian mechanism with per-instance (and global) sensitivity set as the above perturbation bound.
\end{proof}
In the more general smooth-loss case we can simply replace $f^{''}(\hat{\theta}^p; z)\|x\|^2$ with $\|\nabla^2\ell(\hat{\theta}^p;z)\|_F$.

     \begin{algorithm}[H]
        \caption{More adaptive privacy report for \texttt{Obj-Pert}}
        \label{alg:adaptive_report}
        \begin{algorithmic}
            \STATE {\bfseries Input:} $\hat{\theta}^p$ from \texttt{Obj-Pert}, noise parameter $\sigma,\sigma_2,\sigma_3, \sigma_4$; regularization parameter $\lambda$; Hessian $H := \sum_i\nabla^2\ell(\hat{\theta}^p; z_i) + \lambda I_d$, failure probability $\rho$.
            \STATE {\bfseries Output:} Reporting function $\tilde{\epsilon}: (x,y),\rho \rightarrow \R_+^2$.
            \STATE Privately release $\hat{g}^p$ by Algorithm ~\ref{alg:Alg_gp} with parameter $\sigma_2$.
             \STATE Set $\epsilon_2(\cdot)$ according to Theorem $\ref{thm:gp_pdp}$.
             \STATE Set $\overline{g^P}(z) := f'(\cdot) [\hat{g}^P]^Tx + \sigma_2 ||f'(\cdot)x||_2F_{\cN(0,1)}^{-1}(1-\nicefrac{\rho}{2}).$
            \STATE Set $\tau = F_{\lambda_1(\mathrm{GOE}(d))}^{-1}(1-\rho/2)$.
            \vspace{2pt}
            \STATE Privately release $\hat{H}^p$ by Algorithm~\ref{alg:release_H} with parameter $\sigma_3$. 
            \STATE Set $\epsilon_3(\cdot)$ according to Theorem~\ref{thm:pdp_of_H}
            \STATE Privately release $\hat{\lambda}_{\min}^p = \lambda_{\min} + \cN(0,\sigma_4^2)$ (Algorithm~\ref{alg:lambda_min}).
            \STATE Set $\epsilon_4(\cdot)$ according to Theorem~\ref{thm:pdp_of_lambda_min}.
            \STATE Set $\underline{\hat{\lambda}_{\min}^p}:= \max\{\lambda,\hat{\lambda}_{\min}^p - \sigma_4 F_{\cN(0,1)}^{-1}(1-\rho/2)\}$.
            \IF{$\underline{\hat{\lambda}_{\min}^p} \geq 2\tau \sigma_3$}
            \STATE 
            Set $\overline{\mu^p}(x) = \min\Big\{\frac{\underline{\hat{\lambda}_{\min}^p} + \tau \sigma_3}{\underline{\hat{\lambda}_{\min}^p} } x^T (\hat{H}^p)^{-1} x,\frac{\|x\|^2}{\underline{\hat{\lambda}_{\min}^p}}\Big\}.$   \blue{(use the standard estimator)}
            \ELSE  
            \STATE Set $\overline{\mu^p}(x) = \min \Big\{\frac{\underline{\hat{\lambda}_{\min}^p} + 2\tau \sigma_3}{\underline{\hat{\lambda}_{\min}^p} } x^T (\hat{H}^p +  \tau \sigma_3 I_d)^{-1} x,\frac{\|x\|^2}{\underline{\hat{\lambda}_{\min}^p}}\Big\}.$
            \blue{(use the regularized estimator)}
            \ENDIF
            \STATE Set $\overline{\epsilon_1^p}(\cdot):= 
            \left|-\log \big(1 - f''(\cdot)\overline{\mu^P}(x) \big)\right| + \frac{||f'(\cdot)x||_2^2}{2\sigma^2}  +  \frac{|\overline{g^P}(z)|}{\sigma^2} $.
            \STATE Return the ``privacy report'' function  $\tilde{\epsilon} = (\overline{\epsilon_1^p}, \epsilon_2 + \epsilon_3 + \epsilon_4)$, i.e., the \emph{ex-post} pDP of Algorithm~\ref{alg:Alg1} and the pDP of Algorithm~\ref{alg:adaptive_report} (i.e., overhead).
            \vspace*{1mm}
        \end{algorithmic}
    \end{algorithm}

This algorithm allows any choice of $\lambda$ to be used in ObjPert, so that the privacy report is non-intrusive and can be attached to an existing workflow without changing the main algorithm at all. The following proposition shows that $\overline{\mu}^p(x)$ is always a valid upper bound of the leverage score $\mu(x)$ and it is accurate if $\lambda_{\min}$ is large (from either the Hessian or the regularization).

     \begin{proposition}[Uniform multiplicative approximation]\label{prop:utility_H_regularized}
 Let $\underline{\hat{\lambda}_{\min}^p}$ and $\hat{H}^P$ be constructed as in Algorithm~\ref{alg:adaptive_report}.
 Then with  probability $1-2\rho$, 
 $$
\lambda_{\min} - \sigma_4F_{\cN(0,1)}^{-1}(1-\rho/2)\leq \hat{\lambda}_{\min}^p \leq \lambda_{\min} + \sigma_4F_{\cN(0,1)}^{-1}(1-\rho/2)
 $$
and for all $x\in \R^d$ simultaneously, the regularized estimator obeys that
     $$ x^T(\hat{H}^P + \tau \sigma_3 I_d)^{-1}x \leq \mu(x) \leq \frac{\underline{\hat{\lambda}_{\min}^p} + 2\tau\sigma_3}{\underline{\hat{\lambda}_{\min}^p}} x^T(\hat{H}^P + \tau\sigma_3 I_d)^{-1}x. $$
Moreover, under the same high-probability event, if $\underline{\hat{\lambda}_{\min}^p}\geq 2\tau\sigma_3$, then the standard estimator obeys that
$$
\frac{\underline{\hat{\lambda}_{\min}^p} - \tau\sigma_3}{\underline{\hat{\lambda}_{\min}^p}} x^T(\hat{H}^P)^{-1} x \leq x^TH^{-1}x \leq \frac{\underline{\hat{\lambda}_{\min}^p} + \tau\sigma_3}{\underline{\hat{\lambda}_{\min}^p}}  x^T(\hat{H}^P)^{-1} x.
$$
 \end{proposition}
 \begin{proof}
By Lemma~\ref{lem:lambda_GOE_exact}, if we choose $\tau = F_{\lambda_1(\mathrm{GOE}(d))}^{-1}(1-\rho/2)$, then with probability $1-\rho$, the GOE noise matrix $G$ satisfies that  
  $\|G\|_2\prec \tau$, the following holds: $-\tau I_d \prec G\prec \tau I_d$. 
 
 Next, by the definition of Gaussian CDF, with probability $1-\rho$, $$\lambda_{\min} - \sigma_4F_{\cN(0,1)}^{-1}(1-\rho/2) \leq \hat{\lambda}_{\min}^p \leq \lambda_{\min} + \sigma_4 F_{\cN(0,1)}^{-1}(1-\rho/2)$$ which implies that 
 $\lambda_{\min} \geq \underline{\hat{\lambda}_{\min}^p}$, i.e.,
 $$
 H - \underline{\hat{\lambda}_{\min}^p} I_d \succ 0
 $$

  Therefore with probability $1-2\rho$,
  \begin{align*}
H&\prec H + G + \tau \sigma_3 I_d \prec H + 2\tau \sigma_3 I = H - \underline{\hat{\lambda}_{\min}^p} I_d + \underline{\hat{\lambda}_{\min}^p} I_d + 2\tau\sigma_3 I_d \\
&\prec \frac{\underline{\hat{\lambda}_{\min}^p} + 2\tau\sigma_3}{\underline{\hat{\lambda}_{\min}^p}}(H - \underline{\hat{\lambda}_{\min}^p} I_d + \underline{\hat{\lambda}_{\min}^p} I_d) = \frac{\underline{\hat{\lambda}_{\min}^p} + 2\tau\sigma_3}{\underline{\hat{\lambda}_{\min}^p}} H,
\end{align*} 
where the first semidefinite inequality uses that $H - \underline{\hat{\lambda}_{\min}^p} I_d$ is positive semi-definite.

Taking the inverse on both sides, we get
$$
\frac{\underline{\hat{\lambda}_{\min}^p}}{\underline{\hat{\lambda}_{\min}^p} + 2\tau\sigma_3}H^{-1} \prec (\hat{H}^P + \tau \sigma_3 I_d)^{-1} \prec H^{-1}.
$$
Thus for all $x\in \R^d$,
$x^T(\hat{H}^P + \tau \sigma_3 I_d)^{-1}x \leq x^T H^{-1}x \leq \frac{\underline{\hat{\lambda}_{\min}^p} + 2\tau \sigma_3}{\underline{\hat{\lambda}_{\min}^p}} x^T(\hat{H}^P + \tau \sigma_3 I_d)^{-1}x,$
which finishes the proof for the regularized estimator.

Now we turn to the standard (unregularized) estimator. Under the same high-probability event:
  \begin{align*}
H + G\prec& H + \tau \sigma_3 I = H - \underline{\hat{\lambda}_{\min}^p} I_d + \underline{\hat{\lambda}_{\min}^p} I_d + \tau\sigma_3 I_d \\
&\prec \frac{\underline{\hat{\lambda}_{\min}^p} + \tau\sigma_3}{\underline{\hat{\lambda}_{\min}^p}}(H - \underline{\hat{\lambda}_{\min}^p} I_d + \underline{\hat{\lambda}_{\min}^p} I_d) = \frac{\underline{\hat{\lambda}_{\min}^p} + \tau\sigma_3}{\underline{\hat{\lambda}_{\min}^p}} H.
\end{align*} 
Similarly,
\begin{align*}
   H + G\succ& H- \tau\sigma_3 I_d \succ H - \underline{\hat{\lambda}_{\min}^p} I_d + \underline{\hat{\lambda}_{\min}^p} I_d - \tau\sigma_3 I_d \\ \succ&\frac{\underline{\hat{\lambda}_{\min}^p} - \tau\sigma_3}{\underline{\hat{\lambda}_{\min}^p}}(H - \underline{\hat{\lambda}_{\min}^p} I_d + \underline{\hat{\lambda}_{\min}^p} I_d) = \frac{\underline{\hat{\lambda}_{\min}^p} - \tau\sigma_3}{\underline{\hat{\lambda}_{\min}^p}} H.
\end{align*}
Together the above two inequalities give 
$$
\frac{\underline{\hat{\lambda}_{\min}^p} - \tau\sigma_3}{\underline{\hat{\lambda}_{\min}^p}} H \prec H+G \prec \frac{\underline{\hat{\lambda}_{\min}^p} + \tau\sigma_3}{\underline{\hat{\lambda}_{\min}^p}} H.
$$
Take the inverse on both sides we get 
$$
\frac{\underline{\hat{\lambda}_{\min}^p}}{\underline{\hat{\lambda}_{\min}^p} + \tau\sigma_3}H^{-1} \prec (\hat{H}^P)^{-1} \prec \frac{\underline{\hat{\lambda}_{\min}^p}}{\underline{\hat{\lambda}_{\min}^p} - \tau\sigma_3} H^{-1},
$$
which implies that for all $x\in \R^d$, 
$
\frac{\underline{\hat{\lambda}_{\min}^p} - \tau\sigma_3}{\underline{\hat{\lambda}_{\min}^p}} x^T(\hat{H}^P)^{-1} x \leq x^TH^{-1}x \leq \frac{\underline{\hat{\lambda}_{\min}^p} + \tau\sigma_3}{\underline{\hat{\lambda}_{\min}^p}}  x^T(\hat{H}^P)^{-1} x
$
as stated in the proposition.
\end{proof}

The privacy (DP and pDP) of Algorithm~\ref{alg:adaptive_report} is a composition of the stated results in Theorem~\ref{thm:master_thm_reporting} with the the privacy guarantees stated in Theorem~\ref{thm:pdp_of_lambda_min}. Observe that if we choose $\sigma_3 = \sigma_1$ then the additional DP and pDP losses are smaller than those of the main algorithm, i.e., we have a constant overhead in terms of the privacy loss. 

The next theorem shows that when $\lambda_{\min}(H) \rightarrow +\infty$ as the number of data points $n\rightarrow +\infty$, we could improve the leverage score part of the pDP losses from a multiplicative factor of $12$ to $1 + o(1)$.

\begin{theorem}[Utility of Adaptive privacy report.]\label{thm:utility_adaptive_reporting}
Assume $\lambda_{\min}(H) \geq \max\{2\beta, 2\tau\sigma_3\}$.
There is a universal constant $0<C\leq 4\tau \sigma_3 + 2\beta$ such that for a fixed $z \in \cX\times \cY$, and all $\rho > 0$, the privately released privacy report $\overline{\epsilon_1^P}(\cdot)$ from Algorithm~\ref{alg:adaptive_report} obeys that 
$$\epsilon_1(\cdot) \leq \overline{\epsilon_1^P}(\cdot) \leq (1 + \frac{C}{\lambda_{\min}})\epsilon_1(\cdot) + \frac{|f'(\cdot)|\|x\|}{\sigma_2}\sqrt{2\log(2/\rho)}$$
with probability $1-3\rho$ where $\epsilon_1$ is the expression from Theorem~\ref{thm:pdp_alg1_general}.
\end{theorem}
\begin{proof}[Proof of Theorem~\ref{thm:utility_adaptive_reporting}]
Similar to the proof of Theorem~\ref{thm:master_thm_reporting}, it suffices to consider the approximation of the first term when we replace $\mu$ with $\overline{ \mu^p}$. First of all, by a union bound, the high probability event in Proposition~\ref{prop:utility_H_regularized} and the high probability event in Theorem~\ref{thm:data_indep_third} (to bound the third term in the \emph{ex-post} pDP of \texttt{ObjPert}) holds simultaneously with probability at least $1-3\rho$. The remainder of the proof conditions on this event. 

Observe that it suffices to construct a multiplicative approximation bound for the first term $\log(1+f''(\cdot)\mu)$ or $-\log(1-f''(\cdot)\mu)$.

    By our assumption that $\lambda > 2\beta$, as well as the pointwise minimum in the construction of $\overline{\mu^p}$ from Algorithm~\ref{alg:adaptive_report}, we know that $\overline{\mu^p}\leq 1/2$ and $\log(1-f''(\cdot)\overline{\mu^p})$ is well-defined.
    
    Using the fact that for all $a\geq -1$, 
    $
    \frac{a}{1+a} \leq \log(1+a) \leq a,
    $
    we will now derive the multiplicative approximation for both $\log(1+f''(\cdot)\mu)$ or $-\log(1-f''(\cdot)\mu)$ using the plug-ins: $\log(1+f''(\cdot)\overline{\mu^p})$ or $-\log(1-f''(\cdot)\overline{\mu^p})$.
    
    
    For brevity, in the subsequent derivation we will be using $a$ to denote $f''(\cdot)\mu(x)$ and $\hat{a}$ to denote $f''(x^T\hat{\theta}^p; y) \overline{\mu^p}(x)$.
    
    Thus 
    \begin{align*}
    \log(1+a) \leq& \log(1+\hat{a}) \leq \hat{a} \leq(1 + \frac{2\tau\sigma_3}{\underline{\hat{\lambda}_{\min}^p}})a \leq (1 + \frac{2\tau\sigma_3}{\underline{\hat{\lambda}_{\min}^p}})(1+a)\log(1+a) \\
    \leq& (1 + \frac{4\tau\sigma_3}{\lambda_{\min}})(1+\frac{\beta}{\lambda_{\min}})\log(1+a) \leq (1 + \frac{C}{\lambda_{\min}}) \log(1+a)
    \end{align*}
    where $C$ can be taken as $4\tau \sigma_3 + 2\beta$, by our assumption on $\lambda_{\min}$ and a high probability bound under which $\underline{\hat{\lambda}_{\min}^p}\geq \lambda_{\min}/2$.
    
    Similarly,
    \begin{align*}
    -\log(1-a)\leq  \frac{a}{1-a} \leq \frac{\hat{a}}{1-a} \leq \frac{(1 + \frac{2\tau\sigma_3}{\underline{\hat{\lambda}_{\min}^p}})a}{1-a} \leq \frac{(1 + \frac{2\tau\sigma_3}{\underline{\hat{\lambda}_{\min}^p}}) }{1-\frac{\beta}{\lambda_{\min}}} ( -\log(1-a))
    \end{align*}
    where 
    $$
    \frac{(1 + \frac{2\tau\sigma_3}{\underline{\hat{\lambda}_{\min}^p}}) }{1-\frac{\beta}{\lambda_{\min}}} = 1 +  \frac{2\tau\sigma_3}{\underline{\hat{\lambda}_{\min}^p}} + \frac{\beta/\lambda_{\min}}{1-\beta/\lambda_{\min}}\leq 1 + \frac{4\tau\sigma_3 +2\beta}{\lambda_{\min}}    $$
    under our assumption for $\lambda_{\min}$, $\beta$.  The additive error term in the third term follows from the same bound as in the non-adaptive result without any changes.

    The version for the standard (non-regularized) version is similar and is left as an exercise.
\end{proof}

\subsection{Dataset-Dependent Privacy report for general smooth learning problems}
So far, we have focused on generalized linear losses. Most of our results can be extended to general smooth learning problems.  

For the third term in the pDP bound of Theorem~\ref{thm:master_thm_reporting}, the challenge is that the two vectors are now nontrivially coupled with each other via $\hat{\theta}^p$.  For this reason we propose to privately release the gradient at $\hat{\theta}^p$, which helps to decouple the dependence and allow a tighter approximation at a small cost of accuracy and additional privacy budget.


For convenience, we will denote $g = \nabla J(\hat{\theta}^P; D)^T \nabla \ell(\hat{\theta}^P; z)$. Below, we present an algorithm that outputs $g^P$ (a private approximation of $g$) as well as the additional privacy cost $\epsilon_4(\cdot)$ of outputting $g^P$.

 \begin{algorithm}[H]
	\caption{Release $g^P$, a private approximation of $g = \nabla J(\hat{\theta}^P; D)^T \nabla \ell(\hat{\theta}^P; z)$}
	\label{alg:Alg_gp}
	\begin{algorithmic}
		\STATE {\bfseries Input:} Dataset $D$, privatized output $\hat{\theta}^P$, noise parameter $\sigma_2$, linear loss function $L(\theta; D) = \sum_i \ell(\theta; z_i)$, regularization parameter $\lambda$, convex and twice-differentiable regularizer $r$.
		\STATE {\bfseries Output:} $g^P(\cdot), \epsilon_2(\cdot)$.
		\STATE Construct noise vector $e \sim \mathcal{N}(0, \sigma_2^2 I)$.
		\STATE Set $J^P := \nabla L(\hat{\theta}^P; D) + \nabla r(\theta) + \lambda \hat{\theta}^P + e$.
		\STATE Set $g^P(\cdot)$ s.t. $g^P(z) = (J^P)^T \nabla \ell_z(\hat{\theta}^P; z)$.
		\STATE Set $\epsilon_2(\cdot)$ s.t. $\epsilon_2(z)= \frac{\|\nabla \ell(\hat{\theta}^P; z)\|^2}{2\sigma_2^2}  + \frac{\|\nabla \ell(\hat{\theta}^P; z)\|\sqrt{2\log(2/\delta)}}{\sigma_2}$.
	\end{algorithmic}
\end{algorithm}

\begin{theorem} 
   	\label{thm:gp_pdp}
   	Let $\hat{\theta}^P$ be fixed, Algorithm~\ref{alg:Alg_gp} satisfies
   	\begin{enumerate}
   		\itemsep0em
   			\item $(\epsilon_2(D,D_{\pm z}),\delta)$-pDP, with
   			$$\epsilon_2(D,D_{\pm z})=\frac{\|\nabla \ell(\hat{\theta}^P;z)\|^2}{2\sigma_2^2}  + \frac{\|\nabla \ell(\hat{\theta}^P;z)\|\sqrt{2\log(1/\delta)}}{\sigma_2}.   $$
   		   		\item $\epsilon_2(o,D,D_{\pm z})$-ex post pDP with probability $1-\rho$,
   		   		$$
   		   		\epsilon_2(o,D,D_{\pm z}) = \frac{\|\nabla \ell(\hat{\theta}^P;z)\|^2}{2\sigma_2^2}  + \frac{\|\nabla \ell(\hat{\theta}^P;z)\|\sqrt{2\log(2/\rho)}}{\sigma_2}.
   		   		$$
   		   		\end{enumerate}
\end{theorem}
\begin{proof}
This is a Gaussian mechanism and the proof follows from Corollary~\ref{cor:highprob_pdp}.
\end{proof}
The theorem avoids an additional dependence in $d$ from the $\ell_1$-norm $\|\nabla \ell(\hat{\theta}^p; z)\|_1$ in the dataset-independent bound.

We remark that Algorithm~\ref{alg:Alg_gp}'s pDP loss is dataset-independent and if we choose $\sigma_2 = \sigma_1$, the pDP losses for running Algorithm~\ref{alg:Alg_gp} are on the same order as those of the main algorithm. Thus the additional overhead is on the same order and no recursive privacy reporting is needed.

For the first term, our release of $H$ and $\lambda_{\min}$ extends without any changes to the more general case.  The estimator of the leverage score needs to be modified accordingly. 
\yw{Add the plug-in estimator that replaces $H$ with $\hat{H}^p$ in the general case here.}

We defer the analysis of how accurately this estimator approximates the first term of $\epsilon_1(\cdot)$ to a longer version of the paper.


\subsection{Uniform Privacy Report and Privacy Calibration}


The ``privacy report'' algorithm (Algorithm~\ref{alg:privacy_report}) that we presented in the main paper and the ``adaptive privacy report'' (Algorithm~\ref{alg:adaptive_report} is straightforward and omitted. focus on releasing a reporting function $\tilde{\epsilon}$ that is accurate with high probability for every fixed input. 

Sometimes there is a need to ensure that with high probability, $\tilde{\epsilon}$ is accurate \emph{simultaneously} for all $z_1,...,z_n$ in the dataset, or even for all $z\in \cZ$ for a data domain $\cZ$.  The following theorem shows that this is possible at a mild additional cost in the accuracy. These results are stated for Algorithm~\ref{alg:privacy_report}), but extensions to that of Algorithm~\ref{alg:adaptive_report} is straightforward and thus omitted.
    \begin{proposition}[Uniform privacy report]
       \label{prop:uniform_privacy_report}
    With probability $1-2\rho$,  \emph{simultaneously} for all $n$ users in the dataset, the output of Algorithm~\ref{alg:privacy_report} obeys that  $\epsilon_1(\hat{\theta}^p, D, D_{\pm z}) \leq \overline{\epsilon_1^P}(\hat{\theta}^p, z) \leq 12 \epsilon_1(\hat{\theta}^p, D, D_{\pm z}) + \frac{|f'(\cdot)|\|x\|}{\sigma_2}\sqrt{2\log(n/\rho)}$.
    
    If we, instead, use the data-independent bound $\frac{|f'(x^T\hat{\theta}^p; y)|\|x\|_1\sqrt{2\log(2d/\rho)}}{\sigma} $ to replace the third-term in $\overline{\epsilon^P_1}(\cdot)$, then with probability $1-2\rho$, \emph{simultaneously} for all $x\in\cX$, the ex-post pDP report  $\overline{\epsilon_1^P}$ from Algorithm~\ref{alg:privacy_report} satisfies that
    $$\epsilon_1(\cdot) \leq \overline{\epsilon_1^P}(z,\hat{\theta}^p) \leq 12 \epsilon_1(\cdot) + \frac{|f'(\cdot)|\|x\|_1\sqrt{2\log(2d/\rho)}}{\sigma} .$$ 
    \end{proposition}
  \begin{proof}
    We note that the approximation of $\mu_x$ is uniform for all $x$. It remains to consider a uniform bound for the third term over the randomness of ObjPert. The first statement follows by taking a union bound. The second result is achieved by Holder's inequality,  the concentration of max of i.i.d. Gaussians.
    \end{proof}

Sometimes it is desirable to calibrate the noise-level to a prescribed ``worst-case'' DP parameter $\epsilon,\delta$. The following corollary explains that the additional DP loss and pDP losses when we calibrate Algorithm~\ref{alg:privacy_report} with the same privacy parameter as those in Algorithm~\ref{alg:Alg1} will yield a total DP and pDP that are at most twice as large under an additional condition that $f''\leq f'$.
      \begin{corollary}[The additional privacy cost]
    If we calibrate $\sigma_2$ such that the Algorithm~\ref{alg:privacy_report} satisfies the same $(\epsilon,\delta)$-DP as Algorithm~\ref{alg:Alg1}, i.e., when $\epsilon<1$, we could choose $\sigma_2 = \frac{\rho_{\max}}{\epsilon}\sqrt{2\log(1.25/\delta)}$. 
    Then Algorithm~\ref{alg:privacy_report} satisfies $(\epsilon(\cdot),\delta)$-pDP with
        $$
        \epsilon(\cdot) =\frac{\epsilon^2 (f''(\cdot))^2\|x\|^4}{8 \rho_{\max}^2 \log(1.25/\delta)} + \frac{\epsilon (f''(\cdot))\|x\|^2}{\rho_{\max} \sqrt{2}}.$$
    For those cases when $ \frac{(f''(\cdot))\|x\|^2}{\rho_{\max}}\leq  \frac{|f'(\cdot)|\|x\|}{\beta}$ (which is the case in logistic regression for all $x$ s.t., $\|x\|\leq 1$), the additional overhead in releasing a dataset-dependent pDP is smaller than the ex post pDP bound in Theorem~\ref{thm:pdp_alg1_general}.
    \end{corollary}


\section{Improved ``Analyze Gauss'' with Gaussian Orthogonal Ensembles}\label{app:analyze_gauss}


In this section we propose a differentially private mechanism that releases a matrix $H$ when 
$$
H = \sum_{i=1}^n H_x
$$
where $H_x \in \R^{d\times d}$ is a symmetric matrix computed from individual data point $x$.

Examples of this include  
\begin{enumerate}
\item (unnormalized / uncentered) sample covariance  $H_x = xx^T$
\item Empirical Fisher information  $H_x = \nabla \ell(\theta;x) \nabla \ell(\theta;x)T$ where $\ell$ is the log-likelihood and $\theta$ is the true parameter;
\item Hessian of a generalized linear loss function  $H_x = f''(x,\theta) xx^T$.
\item Hessian of a smooth loss function
$H_x = \nabla^2 \ell(x,\theta)$.
\end{enumerate}
In the first three cases $H_x$ is a rank-1 matrix and our use case in this paper is the third and fourth example. 
Throughout this section we assume
$\|H_x\|_F\leq \beta$ for all $x\in \cX$.

The mechanism we propose is a variant of ``Analyze-Gauss'' \citep{dwork2014analyze} but it reduces the required variance of the added noise by a factor of $2$ in almost all coordinates hence resulting in higher utility.

The standard ``Analyze-Gauss'' leverages the symmetry of $H$ and uses the standard Gaussian mechanism to release the upper triangular region (including the diagonal) of the matrix $H$ with an $\ell_2$-sensitivity upper bound:
$$
\|\textrm{UpperTriangle}(H) -  \textrm{UpperTriangle}(H')\|_2 \leq \|H_x\|_F \leq \beta.
$$
where $\textrm{UpperTriangle}(H)\in \R^{d^2/2 + d/2}$ is the vector that enumerates the elements of the upper-triangular region of $H$. The resulting Gaussian noise is distributed i.i.d as $\cN(0,\sigma_3^2)$ and it satisfies $(\epsilon,\delta)$-DP with 
$$\epsilon = \frac{\beta^2}{2\sigma_3^2} + \frac{\beta \sqrt{2\log(1/\delta)}}{\sigma_3}.$$

The alternative that we propose also adds a symmetric noise but doubles the variance on the diagonal elements.
    \begin{algorithm}[H]
        \caption{ Release $H$ (a natural variant of ``Analyze-Gauss'') }
        \label{alg:release_H}
        \begin{algorithmic}
            \STATE{{\bfseries Input:} Dataset $D$, noise parameter $\sigma_3$, $H  = \sum_{i=1}^n \nabla^2\ell(z_i, \hat{\theta}^P) + \lambda I_d$.}
            \STATE{{\bfseries Output:} $\hat{H}^P$.}
           \STATE{Draw a  Gaussian random matrix $Z\in \R^{d\times d}$ with $Z_{i,j}\sim \cN(0,\sigma_3^2)$ independently.}
           \STATE{Output $\hat{H}^P = H + \frac{1}{\sqrt{2}}(Z + Z^T)$.}
        \end{algorithmic}
    \end{algorithm}

The symmetric random matrix $\frac{1}{\sqrt{2}}(Z + Z^T)$ is known as the Gaussian Orthogonal Ensemble (GOE) and well-studied in the random matrix theory. We will first show this this mechanism obeys DP and pDP.

\begin{theorem}\label{thm:pdp_of_H}
    Algorithm~\ref{alg:release_H} satisfies pDP with 
    $$
    \epsilon(\cdot) = \frac{
    \|H_x\|_F^2
    }{4\sigma_3^2} +  \frac{
    \|H_x\|_F
    \sqrt{2\log(1/\delta)}}{\sqrt{2}\sigma_3},
    $$
    and $\hat{H}^p$ satisfies ex post pDP of the same $\epsilon$ with probability $1-2\delta$.
    If in addition 
    $\sup_{x\in \cX}\|H_x\|_F\leq \beta$ then, Algorithm~\ref{alg:release_H} satisfies $(\epsilon,\delta)$-DP with
    $$
    \epsilon \leq \frac{\beta^2}{4\sigma_3^2} +  \frac{\beta\sqrt{2\log(1/\delta)}}{\sqrt{2}\sigma_3}.
    $$
\end{theorem}
\noindent\textbf{Improvements over ``Analyze Gauss''.} Notice that if we choose $\sigma_3$ to be $1/\sqrt{2}$ of the noise scale with used in the standard ``Analyze Gauss'', we will be adding the same amount of noise on the diagonal, achieve the same DP and pDP bounds, while adding noise with only half the variance in the off-diagonal elements. The idea is to add noise with respect to the natural geometry of the sensitivity, as we illustrate in the proof.
\begin{proof}
Algorithm~\ref{alg:release_H} is equivalent to releasing the vector 
$
[f_1, f_2]
$
using a standard Gaussian mechanism with $\cN(0,\sigma_3^2 I_{\frac{d^2}{2}+d/2})$, where $f_1\in\R^d$ is the diagonal of $H/\sqrt{2}$ and $f_2\in\R^{(d^2-d)/2}$ is the vectorized  the strict upper triangular part of $H$.

The per-instance $\ell_2$-sensitivity of $[f_1,f_2]$ is 
\begin{align*}
\|\Delta_x\|_2 &= \sqrt{\sum_{1\leq i<j\leq d}H_{x}[i,j]^2 + \sum_{k=1^d} H_x[k,k]^2 (1/\sqrt{2})^2} \\
&= \sqrt{\frac{1}{2}\left(\sum_{1\leq i<j\leq d} H_{x}[i,j]^2 +\sum_{1\leq j<i\leq d} H_{x}[i,j]^2 + \sum_{k=1^d} H_x[k,k]^2\right)} \\
&= \frac{1}{\sqrt{2}}\|H_x\|_F
\end{align*}

The result then follows from an application of the pDP computation of the Gaussian mechanism. 
\end{proof}

\subsection{Exact statistical inference with the Gaussian Orthogonal Ensemble}
Besides a constant improvement in the required noise, another major advantage of using the Gaussian Orthogonal Ensemble is that we know the exact distribution of its eigenvalues \citep{chiani2014distribution} which makes statistical inference, e.g., constructing confidence intervals, easy and constant-tight.

\begin{lemma}[Largest singular value of Gaussian random matrix {\citep[Equation (2.4)]{rudelson2010non}}]
Let $A\in \R^{d\times d}$ be a random matrix with i.i.d. $\sigma^2$-subgaussian entries, then there exists universal constants $C,c$ such that for all $t>0$
$$
\P[ s_{\max}(A) \geq (2+t)\sqrt{d \sigma^2}] \leq C e^{-c d t^{3/2}}.
$$
i.e., with probability $1-\delta$
$$
\|A\|_2 \leq \left(2 + \big(\frac{(\log(C/\delta))}{c d}\big)^{2/3}\right)\sqrt{d\sigma^2}.
$$
\end{lemma}

Notice that the symmetric matrix, i.e., Gaussian orthogonal ensemble is identically distributed to $\frac{1}{\sqrt{2}}(Z + Z^T)$ where $Z$ is a iid Gaussian random matrix, thus by triangular inequality, we have
\begin{lemma}[Largest eigenvalue of Gaussian orthogonal ensemble]\label{lem:lambda_GOE}
Let $A$ be a Gaussian orthogonal ensemble (i.e., a symmetric random matrix with $\cN(0,\sigma^2)$ on the off-diagonal and $\cN(0,2\sigma^2)$ on the diagonal), with probability $1-\delta$,
$$\|A\|_2 \leq \sqrt{2}\left(2 + \big(\frac{(\log(C/\delta))}{c d}\big)^{2/3}\right)\sqrt{d\sigma^2}.$$
\end{lemma}
\begin{proof}
    The proof follows from triangular inequality of the spectral norm.
\end{proof}

The above bound is asymptotic and we will use it for deriving the theoretical results. For practical computation, 
the the exact formula of the CDF of the largest eigenvalue of GOE matrices is given by \citep[Theorem 2]{chiani2014distribution}. We could use this to bound the spectral norm of the noise added to Algorithm~\ref{alg:release_H}.
\begin{lemma}\label{lem:lambda_GOE_exact}
Let $A$ be described as in Lemma~\ref{lem:lambda_GOE}. 
$$\|A\|_2 \leq \sigma F_{\lambda_1\text{ of GOE}}^{-1}(1-\rho/2)$$
where $F_{\lambda_1\text{ of GOE}}$ is the CDF of the largest eigenvalue of the standard GOE matrix with constructed by $\frac{1}{\sqrt{2}} (Z+Z^T)$ where each element of matrix $Z$ is drawn i.i.d. from a standard gaussian.
\end{lemma}
\begin{proof}
    Notice that the GOE matrix is symmetric, so the largest eigenvalue $\lambda_1$ and the negative of the smallest eigenvalue $-\lambda_d$ are identically distributed.  Thus the operator norm $\|A\|_2 \leq \max\{ |\lambda_1|, |\lambda_d|\} \leq F_{\lambda_1\text{ of GOE}}^{-1}(1-\rho/2)$ with probability $1-\rho$.
\end{proof}
 
\noindent\textbf{Numerical computation:}
 \citet[Theorem 2]{chiani2014distribution} characterized the distribution of $\lambda_1$ and provided an exact analytical formula with stable numerical implementation to compute $F_{\lambda_1\text{ of GOE}}$. Thus $F_{\lambda_1\text{ of GOE}}^{-1}$ can be evaluated using a binary search.
 
 Using the Mathematica implementation provided by ~\citep{chiani2014distribution}, we find that $F_{\lambda_1\text{ of GOE(50)}}^{-1}(1 - \rho/2) = 12$ for $\rho = 8.465 \times 10^{-6}$. Therefore in our experiments with $d = 50$, we choose $\tau \approx 12$.
 
 \yw{Explain how we got that $1.7\sqrt{d}$ constant here (what is the corresponding $\rho$ when we choose $1.7$?). The author has also provided a Mathematica implementation. I don't have one installed but I think UCSB has it for free for students. I think it will be an ease of mind if we can try running the authors' implementation to figure out and comment here that we got this number not just from the figure but also from the author's provided code. You may find the code here \url{https://sites.google.com/site/marcochianigroup/articles}}
 
 \yw{This is of a slightly lower-priority, but I think we should provide an implementation of this in python and make it available.  A general inference tool for the GOE-Analyze-Gauss is very useful.}

\section{Omitted Proofs}
\label{app:proofs}


With the two technical components presented, we are now ready to present the detailed proofs of our main results:  Theorem~\ref{thm:pdp_alg1_general} and Theorem~\ref{thm:master_thm_reporting}.


\subsection{Proofs for the pDP analysis of objective perturbation}


\begin{proof}[Proof of Theorem ~\ref{thm:pdp_alg1_general}]

We calculate the \emph{ex-post} pDP loss of Algorithm ~\ref{alg:Alg1} as follows. Consider the perturbed objective function:
\begin{align}
\label{arg_min}
    \hat{\theta}^P = \argmin\limits_{\theta \in \mathbb{R}^d} \hat{\mathcal{L}}(\theta;D) + r(\theta) + \frac{\lambda}{2}||\theta||_2^2 + b^T\theta.
\end{align}

Let $\mathcal{A}$ be the algorithm that outputs $\hat{\theta}^P$ as stated in \ref{arg_min}. The \emph{ex-post} per-instance privacy loss (with the abuse of notation discussed in Section ~\ref{notation}) is then given by
\begin{align*}
    \epsilon_1(\theta, D, D_{\pm z}) &= \text{max}\left(\log \dfrac{\text{Pr}\big(\mathcal{A}(D) = \hat{\theta}^P \big)}{\text{Pr}\big(\mathcal{A}(D_{\pm z}) = \hat{\theta}^P \big)}\:,\: \log\dfrac{\text{Pr}\big(\mathcal{A}(D_{\pm z}) = \hat{\theta}^P\big)}{\text{Pr}\big(\mathcal{A}(D) = \hat{\theta}^P\big)}\right),
\end{align*}
Note that this characterization of \emph{ex-post} per-instance DP is equivalent to that stated in Definition ~\ref{def:expost_pdp}, since switching the numerator and denominator of a log ratio is the same as flipping its sign.

Since we can't easily calculate the distribution of $\hat{\theta}^P$, we will instead use the bijection between the output $\hat{\theta}^P$ and the noise vector $b$ (observed in ~\cite{chaudhuri2011differentially}) to rewrite the log probability ratio more cleanly.

First-order conditions applied to (~\ref{arg_min}) tell us that \begin{align} \label{b_J_relate}
    b(\hat{\theta}^P;D) &= -\left( \nabla \hat{\mathcal{L}}(\hat{\theta}^P;D)+\nabla r(\hat{\theta}^P) + \lambda \hat{\theta}^P\right).
\end{align}
Then taking the gradient of the noise vector, we have
\begin{align}
    \nabla b(\hat{\theta}^P;D) &= -\left( \nabla^2 \hat{\mathcal{L}}(\hat{\theta}^P;D)+\nabla^2 r(\hat{\theta}^P) + \lambda I_d\right).
\end{align}

Let $b \sim \mathcal{N}(0, \sigma^2I_d)$, and denote $\nu(\cdot)$ as the probability density function of the normal distribution: i.e., the density at $b$ is $\nu(b; \sigma) \propto e^{-\frac{||b||_2^2}{2\sigma^2}}$. Then since the objective function $J(\theta; D)$ is strictly convex in $\theta$ (implying as in ~\cite{chaudhuri2011differentially} that the mapping between $\hat{\theta}^P$ and $b$ is bijective and monotonic), by Lemma ~\ref{lemma:bijection} we can write

\begin{align*}
    \log \dfrac{\text{Pr}\big(\mathcal{A}(D) = \hat{\theta}^P \big)}{\text{Pr}\big(\mathcal{A}(D_{\pm z}) = \hat{\theta}^P\big)} &= \log  \cfrac{\Big|\text{det}\Big(\nabla b(\hat{\theta}^P;D)\Big)\Big|}{\Big|\text{det}\Big(\nabla b(\hat{\theta}^P;D_{\pm z})\Big)\Big|} \cfrac{\nu(b(\hat{\theta}^P;D); \sigma)}{\nu(b(\hat{\theta}^P;D_{\pm z});\sigma)}
    \\
    &= \log  \cfrac{\Big|\text{det}\Big(\nabla b(\hat{\theta}^P;D)\Big)\Big|}{\Big|\text{det}\Big(\nabla b(\hat{\theta}^P;D_{\pm z})\Big)\Big|}\: + \log \cfrac{e^{-\frac{1}{2\sigma^2}||b(\hat{\theta}^P;D)||_2^2}}{e^{-\frac{1}{2\sigma^2}||b(\hat{\theta}^P;D_{\pm z})||_2^2}} \\
     &= \underbrace{\log  \cfrac{\Big|\text{det}\Big(\nabla b(\hat{\theta}^P;D)\Big)\Big|}{\Big|\text{det}\Big(\nabla b(\hat{\theta}^P;D_{\pm z})\Big)\Big|}}_{(*)}\:+\:\underbrace{\frac{1}{2\sigma^2}\Big(||b(\hat{\theta}^P;D_{\pm z})||_2^2 - ||b(\hat{\theta}^P;D)||_2^2\Big)}_{(**)}.
\end{align*}

Dealing first with the term (*), we observe that $\nabla b(\hat{\theta}^P;D_{\pm z}) =  \nabla b(\hat{\theta}^P;D) \mp \nabla^2 \ell(\hat{\theta}^P; z)$. The notation "$\mp$"  means to subtract if $z \notin D$, and add if $z \in D$. Using the eigendecomposition $\nabla^2 \ell(\hat{\theta}^P; z) = \sum_{k = 1}^{d}\lambda_k u_ku_k^T$ and recursively applying the matrix determinant lemma, we have

\begin{align*}
    \Big|\text{det}\big(\nabla b(\hat{\theta}^P;D_{\pm z})\big)\Big| & = \Big|\text{det}\Big(\nabla b(\hat{\theta}^P;D) \mp \nabla^2 \ell(\hat{\theta}^P;z)\Big)\Big| \\
    &= \Big|\text{det}\Big(\nabla b(\hat{\theta}^P;D) \mp  \sum_{k = 1}^{d}\lambda_k u_ku_k^T\Big)\Big|\\
    &= \Big|\text{det}\Big(\nabla b(\hat{\theta}^P;D) \mp \sum_{k = 1}^{d - 1}\lambda_k u_ku_k^T \mp \lambda_d u_du_d^T\Big)\Big|\\
    &= \Big|\text{det}\Big( \nabla b(\hat{\theta}^P; D) \mp \sum\limits_{k = 1}^{d - 1} \lambda_k u_k u_k^T  \Big)\Big| \Big( 1 \mp \lambda_d u_d^T\big(  \nabla b(\hat{\theta}^P; D) \mp \sum\limits_{k = 1}^{d - 1} \lambda_k u_k u_k^T\big)^{-1}u_d\Big) \\
    &=\: \ldots \\
    &= \Big|\text{det}\Big( \nabla b(\hat{\theta}^P; D) \Big)\big| \prod\limits_{j = 1}^d\big(1 \mp \mu_j \big),
\end{align*}
    where $\mu_j =  \lambda_j u_j^T\Big(\nabla b(\hat{\theta}^P; D) \mp \sum_{k = 1}^{j - 1} \lambda_k u_k u_k^T \Big)^{-1}u_j$. Therefore,
    
    \begin{align*}
        (*) &=  \log \dfrac{\Big|\text{det}\big(\nabla b(\hat{\theta}^P;D)\big)\Big|}{  \Big|\text{det}\big(\nabla b(\hat{\theta}^P;D_{\pm z})\big)\Big|} \\
        &=  \log \dfrac{\Big|\text{det}\big(\nabla b(\hat{\theta}^P;D)\big)\Big|}{  \Big|\text{det}\Big( \nabla b(\hat{\theta}^P; D) \Big)\big| \prod\limits_{j = 1}^d\big(1 \mp \mu_j \big)} \\
        &= \log \dfrac{1}{\prod\limits_{j = 1}^d\big(1 \mp \mu_j \big)} \\
        &= -\log \prod\limits_{j = 1}^d\big(1 \mp \mu_j \big).
    \end{align*}
    
    We'll handle the second term (**) next. We have that
    
    \begin{align*}
        (**) &= \frac{1}{2\sigma^2}\Big(||b(\hat{\theta}^P; D_{\pm z})||_2^2 - ||b(\hat{\theta}^P; D)||_2^2\Big) \\
        &= \frac{1}{2\sigma^2} \Big[\mp \nabla \ell(\hat{\theta}^P; z)\Big] \Big[ 2 b(\hat{\theta}^P; D) \mp \nabla \ell(\hat{\theta}^P; z) \Big] \\
        &= \pm \frac{1}{\sigma^2} \big[ \nabla J(\hat{\theta}^P; D)^T \nabla \ell(\hat{\theta}^P; z)\big] + \frac{1}{2\sigma^2}||\nabla \ell(\hat{\theta}^P; z)||_2^2.\\
    \end{align*}

The rest of the proof follows from adding together (*) and (**), and taking the absolute value.
\end{proof}

\begin{proof}[Proof of Corollary ~\ref{cor:pdp_alg1_linear}]
By restricting to generalized linear models, we can give a more interpretable pDP result for Algorithm ~\ref{alg:Alg1} with the main difference being a cleaner version of the generalized leverage score. In the case of GLMs, we have that $\nabla \ell(\cdot) = f'(\cdot)x$ and $\nabla^2 \ell(\cdot) = f''(\cdot )xx^T$. So $\ell(\cdot)$ has a rank-one Hessian with only one eigenvalue, and $\log\prod\limits_{j = 1}^d \Big(1 + \mu_j \Big) = \log (1 + \mu(x))$. Here $\mu_j$ is as defined in Theorem ~\ref{thm:pdp_alg1_general} and $\mu$ is defined as in Corollary ~\ref{cor:pdp_alg1_linear}.
\end{proof}

\begin{proof}[Proof of Theorem \ref{thm:data_indep_first}]
Using the eigendecomposition $\nabla^2 \ell(\hat{\theta}^P; z) = \sum_{k = 1}^d \lambda_k u_k u_k^T$, for $0 \leq j \leq d$ we have that
    \[ \mu_j(x) = \begin{cases} 
          \lambda_j u_j^T \Big( -\nabla^2 L(\hat{\theta}^P; D) - \lambda I_d - \nabla^2 r(\hat{\theta}^P) - \sum_{k = 1}^{j - 1}\lambda_k u_k u_k^T\Big)^{-1} u_j& \text{ if } z \notin D \\
           \lambda_j u_j^T \Big( -\sum\limits_{\substack{z_i \in D \\ z_i \neq z}}\nabla^2 \ell(\hat{\theta}^P; z_i) - \lambda I_d - \nabla^2 r(\hat{\theta}^P) - \sum_{k = j}^{d}\lambda_k u_k u_k^T\Big)^{-1} u_j & \text{ if } z \in D. \\
      \end{cases}
    \]
        \[  := \begin{cases} 
          \lambda_j u_j^T H_{+ z}^{-1} u_j& \text{ if } z \notin D \\
           \lambda_j u_j^T H_{-z}^{-1} u_j & \text{ if } z \in D.
       \end{cases}
    \]
The second equality introduces the shorthand $\mu_j(x) := \lambda_j u_j^T H^{-1}_{\pm z} u_j$. Observe that $\nabla^2 \ell(\hat{\theta}^P; z_i), \nabla^2 r(\hat{\theta}^P) \in \mathbb{R}^{d \times d}$ are positive semi-definite, since $\ell(\cdot)$ and $r(\theta)$ by assumption are convex functions with continuous second-order partial derivatives. Since $\nabla^2 \ell(\hat{\theta}^P; z_i)$ is PSD, its eigenvalues are non-negative and so $\lambda_k \geq 0$ for all $0 \leq k \leq d$. Then for any $x \in \mathbb{R}^d$, $x^Tu_k u_k^Tx = (x^Tu_k)^2 \geq 0$. So $u_ku_k^T$ is also PSD, and we then have that $H_{+z} + \lambda I_d$ and $H_{-z} + \lambda I_d$ are both negative semi-definite. Therefore,  $H_{\pm z} \prec - \lambda I_d$ and after taking the inverse, we see that $\mu_j(x) \leq -\frac{\lambda_j}{\lambda} \leq 0$ or equivalently $-\mu_j(x) \geq \frac{\lambda_j}{\lambda} \geq 0$.

For  $-1 < \mu_j(x) \leq 0$, we have that 
\begin{align*}
\big| -\log(1 - \mu_j(x))\big| &= \log(1 + (- \mu_j(x))) \\ &\leq -\mu_j(x) \\ &\leq -\log(1 + \mu_j(x)) \\ &= \big|- \log(1 + \mu_j(x))\big| \\
&\leq - \log(1 - \frac{\lambda_j}{\lambda}).
\end{align*}
The rest of the proof follows from converting the log-product into a sum of logs. For a linear loss function $\ell(\theta; z) = f(x^T \theta ; y)$, the simplified bound can be achieved due to the rank-one Hessian $\nabla^2 \ell(\hat{\theta}^P; z) = f''(x^T \theta; y) xx^T$ whose only eigenvalue is $\lambda_1 = f''(x^T \hat{\theta}^P; y) ||x||_2^2$.
\end{proof}

\begin{proof}[Proof of Theorem \ref{thm:data_indep_third}]
By Holder's inequality, 
\begin{align*}
    \Big|\nabla J(\hat{\theta}^P; D) \nabla \ell(\hat{\theta}^P; z) \Big| &\leq || \nabla J(\hat{\theta}^P) ||_{\infty} || \ell(\hat{\theta}^P; z) ||_{1}.
\end{align*}
Recall from (\ref{b_J_relate}) that $\nabla J(\hat{\theta}^P; D) = -b(\hat{\theta}^P; D)$. Therefore $|| \nabla J(\hat{\theta}^P) ||_{\infty}  = \max\limits_{i \in [d]} |b_i|$, where $b_i \sim \mathcal{N}(0, \sigma^2)$. Applying a union bound and using the standard Gaussian tail bound,
\begin{align*}
    \text{Pr}\left[\max\limits_{i \in [d]} |b_i| \geq t\right] &= \text{Pr}\left[\bigcup_i |b_i| \geq t \right] \\
    &\leq \sum_{i\in [d]} \text{Pr} \big[|b_i| \geq t \big] \\
    & \leq 2de^{-\frac{t^2}{2\sigma^2}}.
\end{align*}
So with probability $1 - \rho$, we have $|| \nabla^J (\hat{\theta}^P; D) ||_{\infty} \leq \sigma \sqrt{2 \log(2d/\rho)}$. The stronger bound for linear loss functions comes from substituting $||\nabla \ell(\hat{\theta}^P)||_1 = f'(x^T \theta; y) ||x||_1$.
\end{proof}

\subsection{Proofs for the Privacy Report in the main paper}


The proof of Theorem~\ref{thm:master_thm_reporting} relies on the following intermediate result. 
\begin{proposition}[Uniform multiplicative approximation]\label{prop:utility_H_large_lambda}
	If $\lambda_{\min}(H) \geq  2 \sigma_2 F_{\lambda_1(\mathrm{GOE}(d))}^{-1}(1-\rho/2)$, then with probability $1-\rho$, for all $x\in \R^d$ simultaneously 
	$$
	\frac{1}{2}  x^T(\hat{H}^P)^{-1}x \leq x^TH^{-1}x \leq \frac{3}{2}x^T(\hat{H}^P)^{-1}x.
	$$
\end{proposition}
\begin{proof}
	By the choice of $\tau = F_{\lambda_1(\mathrm{GOE}(d))}^{-1}(1-\rho/2)$, with probability $1-\rho$, the noise matrix $Z$ from the release of $\hat{H}^P$ satisfies that $\|Z\|_2 \leq\sigma_2 \tau \leq  \lambda_{\min}/2$.
	Thus $-\frac{H}{2} \prec -\frac{\lambda_{\min}}{2} I_d \prec Z \prec \frac{\lambda_{\min}}{2} I_d  \prec \frac{H}{2}$.
	Adding $H$ on both sides
	$$
	\frac{H}{2} \prec H + Z \prec  \frac{3H}{2}
	$$
	which implies that 
	$$
	\frac{2}{3} H^{-1}\leq (H + Z)^{-1} \prec 2 H^{-1}.
	$$
	By definition of semidefinite ordering, for all $x\in\R^d$	
		\begin{align*}
	\frac{2}{3} x^TH^{-1}x \leq x^T(H+Z)^{-1}x \leq  2x^TH^{-1}x.
	\end{align*} 
	In other word,  $\frac{1}{2} \hat{\mu}^p_1(x)\leq \mu_1(x) \leq \frac{3}{2}\hat{\mu}_1^p(x)$.
\end{proof}

\begin{proof}[Proof of Theorem~\ref{thm:master_thm_reporting}] 
    The privacy guarantees (Statement 1-3) follow directly from the pDP analysis in Theorem~\ref{thm:pdp_of_H} that analyzes the release of $H$ by adding a GOE noise matrix and the Gaussian mechanism that releases $g$.
    
    By the result follows from Proposition~\ref{prop:utility_H_large_lambda} we know that with probability $1-\rho$, for all $x$ 
    $$\mu(x) \leq \frac{3}{2}\hat{\mu}^p(x)\leq 3\mu(x)$$
    
    For all $a\geq -1$
    $
    \frac{a}{1+a} \leq \log(1+a) \leq a.
    $
    Recall that $\beta \geq \sup_z \|\nabla^2\ell(\hat{\theta}^p;z)\|2$. By our condition that $\lambda > 2\beta$, as well as the pointwise minimum in the construction of $\overline{\mu^p}$, we have that
    $f''\overline{\mu^p} \leq \frac{1}{2}$ and
    $$\frac{f''\overline{\mu^p}}{2}\leq \max\{\log(1+f''\overline{\mu^p}), -\log(1-f''\overline{\mu^p}\} \leq 2 f''\overline{\mu^p}.$$
    Thus 
    $$
    \log(1+f''\mu) \leq f''\mu \leq f''\overline{\mu^p}\leq 2\log(1+f''\overline{\mu^p}) \leq 2 f''\overline{\mu^p} \leq 3 f''\hat{\mu}^p\leq 6f''\mu \leq 12\log(1+f''\mu),
    $$
    and similarly
    {\small
    $$
    -\log(1-f''\mu) \leq 2f''\mu \leq 2\overline{f''\mu^p}\leq -2\log(1-f''\overline{\mu^p}) \leq 4f''\overline{\mu^p} \leq 6 f''\hat{\mu}^p \leq  12 f''\mu \leq -12\log(1-f''\mu).
    $$
    }
    This concludes the factor $12$ multiplicative approximation in the first term of  $\epsilon_1(\cdot)$. The second term of $\epsilon_1(\cdot)$ does not involve an approximation. The third term of $\epsilon_1(\cdot)$ is random and the bound is off by an additive factor of $\min\{\sigma,\sigma_2\}|f'(\cdot)| \|x\|_2 \sqrt{2\log(2/\rho)}$ --- via the smaller of the data-dependent bound and the data-independent bound, each holds with probability $1-\rho/2$. 
    \end{proof}


\section{pDP Analysis of the Gaussian mechanism}
\label{app:pdp_gaussian_mech}

\begin{theorem}[\emph{ex-post} pDP of Gaussian mechanism]
    \label{pdp_gaussian}
	Let $Q:  \cZ^* \rightarrow \R^d$ be a function of the data. Let $|Q(D_{\pm z}) - Q(D)| \leq \Delta_z$. Then the Gaussian mechanism that releases $ o \sim Q(D)  + \cN(0,  \sigma^2 I_d)$ obeys \emph{ex-post} pDP with
	$$
	\epsilon(o, D, D_{z}) =  \left| \frac{\|\Delta_z\|^2}{2\sigma^2}  - \frac{ \Delta_z^T(o - Q(D))  }{\sigma^2}  \right|.
	$$
\end{theorem}
\begin{proof} 
	We can directly calculate the log-odds ratio:
	\begin{align*}
	&\frac{1}{2\sigma^2} \left(  \|o - Q(D)\|^2 -  \|o - Q(D_{\pm z})\|^2 \right)  \\
	=& \frac{1}{2\sigma^2} \left(  (Q(D_{\pm z})- Q(D))^T(2o - Q(D) - Q(D_{\pm z})) \right)\\
	=&  \frac{1}{2\sigma^2} \left(  \Delta_z^T(2o - 2Q(D) - \Delta_z) \right)\\
	=&  \frac{- \|\Delta_z\|^2}{2\sigma^2}  +  \frac{\Delta_z^T(o - Q(D))}{\sigma^2}.
	\end{align*}
	The proof is complete by taking the absolute value. 
\end{proof}
\begin{corollary}[pDP bound and high-probability ex-post pDP of Gaussian mechanism]\label{cor:highprob_pdp}
Let $\Phi$ be the cumulative distribution function (CDF) of a standard normal random variable. The Gaussian mechanism that releases $ o \sim Q(D)  + \cN(0,  \sigma^2 I_d)$ satisfies dataset independent pDP bound with 
$$
\epsilon(D,D_{\pm z}) \leq \frac{\|\Delta_z\|^2}{2\sigma^2}  + \frac{\|\Delta_z\|\Phi^{-1}(1-\delta)}{\sigma} \leq \frac{\|\Delta_z\|^2}{2\sigma^2}  + \frac{\|\Delta_z\|\Phi^{-1}(1-\delta)}{\sigma}.
$$
Moreoever, with probability at least $1-\rho$ over the distribution of the randomized output $o$, the Gaussian mechanism satisfies obeys the following dataset-independent \emph{ex post} pDP bound
	\begin{equation}\label{eq:pdp_gm}
	\epsilon(o, D, D_{\pm z}) \leq \frac{\|\Delta_z\|^2}{2\sigma^2}  + \frac{\|\Delta_z\|\Phi^{-1}(1-\rho/2)}{\sigma}\leq \frac{\|\Delta_z\|^2}{2\sigma^2}  + \frac{\|\Delta_z\|\sqrt{2\log(2/\rho)}}{\sigma}.
	\end{equation}
\end{corollary}
\begin{proof}
	Since $ o \sim Q(D)  + \cN(0,  \sigma^2 I_d)$, we have 
	$\Delta_z^T(o - Q(D)) \sim  \cN(0, \sigma^2 \|\Delta_z\|^2)$.
	The results of pDP follows from the tailbound of the privacy loss random variable and Lemma~\ref{lem:tailbound2DP}. 
	
	
	For the high-probability bound of the \emph{ex post} pDP, we need to bound both sides of the privacy loss random variable. It suffices to show that the absolute value of the added noise is bounded with a union bound on the two-sided tails, each with probability $1-\rho/2$.
\end{proof}

A tighter pDP bound can be obtained using the analytical Gaussian mechanism~\citep{balle2018improving}. We choose to present the tail bound-based formula above for the interpretability of the results.

\section{Technical Lemmas}
\label{app:lemmas}

\begin{lemma}[Sherman-Morrison-Woodbury Formula]\label{lem:woodbury}
	Let $A,U,C,V$ be matrices of compatible size. Assuming $A,C$ and $C^{-1}+VA^{-1}U$ are all invertible, then
	$$
	(A+ UCV)^{-1} = A^{-1} - A^{-1}U(C^{-1}+VA^{-1}U)^{-1}VA^{-1}.
	$$
\end{lemma}

\begin{lemma}[Determinant of Rank-1 perturbation]\label{lem:determinant}
	For invertible matrix $A$ and vector $c,d$ of compatible dimension
	$$\det(A + cd^T)  = \det(A)(1+d^TA^{-1}c).$$
\end{lemma}


\begin{lemma}[Gaussian tail bound]\label{lem:gaussian_tail}
	Let $X\sim \cN(0,\sigma^2)$. Then 
	$$
	\P(X >\sigma\epsilon) \leq \frac{e^{-\epsilon^2/2}}{\epsilon}.
	$$
	A convenient alternative representation (slightly weaker) is
	$$
	\P(X > \sigma \sqrt{2\log(1/\delta)}) \leq \delta,
	$$
	and
	$$
	\P(|X| > \sigma \sqrt{2\log(2/\delta)}) \leq \delta.
	$$
	for all $\delta >0$.
\end{lemma}

\begin{lemma}[Tail bound to $(\epsilon,\delta)$-DP conversion]\label{lem:tailbound2DP}
	Let $\epsilon(o) = \log(\frac{p(o)}{p'(o)})$ where $p$ and $p'$ are densities of $\theta$. If 
	$$
	\P_p(\epsilon(o) > \epsilon) \leq \delta
	$$
	then for any measurable set $\cS$
	$$
	\P_p(\theta \in \cS)  \leq e^\epsilon \P_{p'}(\theta \in \cS) + \delta.
	$$
	Two useful applications of this result for DP are:
	\begin{enumerate}
		\item if 	$
		\P_p(\epsilon(o) > \epsilon) \leq \delta
		$
		for all pairs of neighboring dataset $D,D'$ such that $p = \cA(D),p' = \cA(D')$ then $\cA$ is $(\epsilon,\delta)$-DP.
		\item 	If $D' = D_{\pm z}$, $p = \cA(D),p' = \cA(D_{\pm z})$  and that  
		$
		\P_p(\epsilon(o) > \epsilon) \leq \delta
		$
		and
		$
		\P_{p'}(-\epsilon(o) < -\epsilon) \leq \delta,
		$
		then $\cA$ satisfies $(\epsilon,\delta)$-pDP for individual $z$ and dataset $D$.
	\end{enumerate}
\end{lemma}
\begin{proof}
	Let $E$ be the event that $|\epsilon(\theta)| > t$, by definition it implies that for any $\tilde{E}\subset E$, $\P_p(\theta \in \tilde{E}) \leq e^{t} \P_{p'}(\theta \in \tilde{E})$.
	Now consider any measurable set $\cS$:
	\begin{align*}
	\P_p(\theta \in \cS)  &=  \P_p(\theta \in \cS \cap E^c) + \P_p(\theta \in \cS \cap E) \\
	&\leq  \P_{p'}(\theta \in \cS \cap E^c) e^t + \P_p(\theta\in E)\leq e^t\P_{p'}(\theta \in \cS) + \delta.
	\end{align*}
	The two applications follow directly from the definitions of $(\epsilon,\delta)$-DP and pDP.
\end{proof}

\begin{lemma}[maximum of subgaussian]\label{lem:max_of_subgaussian}
Let $X_1,...,X_n$ be iid $\sigma^2$-subgaussian random variables.
$$
\P[\max_i X_i \geq \sqrt{2\sigma^2(\log n + t)}] \leq e^{-t}.
$$
\end{lemma}
\begin{proof}
    The proof is by standard subgaussian concentration and union bound.
\end{proof}

\begin{lemma} [Weyl's theorem; Theorem 4.11, p. 204 in \cite{stewart1990matrix}]. Let $A, E$ be given $m \times n$ matrices with $m \geq n$, then
\begin{align}
\max_{i \in [n]} |\sigma_i(A)-\sigma_i(A+E)| \leq \norm{E}_2
\end{align}
\end{lemma}

\begin{lemma}["Change-of-variables" for density functions]
\label{lemma:bijection}
Let $g:\mathbb{R}^d \rightarrow \mathbb{R}^d$ be a bijective and differentiable function, and let $X, Y$ be continuous random variables in $\mathbb{R}^d$ related by the transformation $Y = g(X)$. Then the probability density of $Y$ is
\begin{align*}
    f_Y(y) &= f_X(g^{-1}(y))\left|\text{det}\left[\dfrac{\partial g^{-1}(y)}{\partial y} \right]\right|,
\end{align*}
with $\left[\dfrac{\partial g^{-1}(y)}{\partial y} \right]$ denoting the $d \times d$ Jacobian matrix of the mapping $X = g^{-1}(Y)$.
\end{lemma}

\begin{lemma} \emph{(Billboard lemma)}
Suppose $\mathcal{A}: \mathcal{D} \rightarrow \mathcal{R}$ satisfies $(\epsilon, \delta)$ differential privacy. Consider any set of functions $f_i: \mathcal{D}_i \times \mathcal{R} \rightarrow \mathcal{R}$, where $\mathcal{D}_i$ is the portion of the dataset containing individual $i$'s data. The composition $\{ f_i(\Pi_i D, \mathcal{A}(D))\}$ satisfies $(\epsilon, \delta)$-joint differential privacy, where $\Pi_i: \mathcal{D} \rightarrow \mathcal{D}_i$ is the projection to individual $i$'s data.
\end{lemma}
\end{document}